%% file: Farfalle_Revision_V5.tex
\documentclass[runningheads]{llncs}

\usepackage{multirow}
\usepackage{hhline}
\usepackage{array}
\usepackage{makecell}

\usepackage{graphicx,epsfig}
\usepackage{epstopdf}

\usepackage{standalone}
\usepackage{tikz}
\usetikzlibrary{calc}
\usetikzlibrary{crypto.symbols}
\usepackage{ifthen}

\usetikzlibrary{backgrounds}
\usetikzlibrary{calc}
\usetikzlibrary{decorations.pathreplacing}
\usetikzlibrary{shapes}

\def\circlearrow#1{\draw[arrow] (#1.center) ++(70:0.30cm) arc (70:410:.30cm)}

\usepackage[hidelinks]{hyperref}

\usepackage{float}
\usepackage{tocloft}
\usepackage{amsfonts,amsmath,amssymb}
\usepackage{multirow, verbatim}
\usepackage{shadow,enumerate}
\usepackage{makeidx}  
\usepackage{graphicx}
\usepackage{color}
\def\Z{{\mathbb Z}}

\usepackage{linegoal}
\usepackage{subcaption}
\usepackage{cite}
\usepackage{cleveref}

\usepackage{algorithm}
\usepackage{algpseudocode}
\newcommand{\N}{\mathbb{N}}
\usepackage{booktabs}

\usepackage{mathrsfs}

\def\whitebox{{\hbox{\hskip 1pt
        \vrule height 6pt depth 1.5pt
        \lower 1.5pt\vbox to 7.5pt{\hrule width
                  3.2pt\vfill\hrule width 3.2pt}%
        \vrule height 6pt depth 1.5pt
        \hskip 1pt } }}
\def\qed{\ifhmode\allowbreak\else\nobreak\fi\hfill\quad\nobreak\whitebox\medbreak}

\begin{document}
\title{Quantum Cryptanalysis of Farfalle and (Generalised) Key-Alternating Feistel Networks}

\author{S. Hod\v zi\'c\inst{1,2} \and A. Roy\inst{3} \and E. Andreeva\inst{4}}

\institute{University of Primorska, FAMNIT, Koper, Slovenia, \email{samir.hodzic@famnit.upr.si} \and
SGS Digital Trust Services, Graz, Austria, \email{samir.hodzic@sgs.com} \and
University of Klagenfurt, Austria, \email{arnab.roy@aau.at}\and
Technical University of Vienna, Austria, \email{elena.andreeva@tuwien.ac.at}
}

\date{}
\maketitle

\begin{abstract}
Farfalle is a permutation-based construction for building a pseudorandom function which has been proposed by G. Bertoni et al. in 2017. In this work, we show that by observing suitable inputs to Farfalle, one can derive various constructions of a periodic function with a period that involves a secret key. As this admits the application of Simon's algorithm in the so-called \textbf{Q2} attack model, we further show that in the case when internal rolling function is linear, then the secret key can be extracted under feasible assumptions. Furthermore, using the provided constructions of periodic functions for Farfalle, we show that one can mount forgery attacks on the session-supporting mode for authenticated encryption (Farfalle-SAE) and the synthetic initial value AE mode (Farfalle-SIV). In addition, as the wide block cipher mode Farfalle-WBC is a 4-round Feistel scheme, a quantum distinguisher is constructed in the case when input branches are containing at last two blocks, where length of one block corresponds to the size of a permutation employed in Farfalle (a similar attack can be mounted to Farfalle-WBC-AE). And finally, we consider the problem of extracting a secret round key out of different periods obtained from a (Generalized)  Feistel scheme (GFN), which has not been addressed in any of the previous works which consider the application of Simon's (or Simon-Grover) algorithm to round reduced versions of GFNs. In this part, we assume that the key is added to an input of an inner function utilized in the round function of a given GFN. By applying two different interpolation formulas, we show that one can extract the round key by utilizing amount of different periods which is closely related to the polynomial/algebraic degree of underlying inner function. Our methods can be seen as an extension of existing quantum attacks on key-alternating GFNs based on Simon's or Simon-Grover algorithms.

\end{abstract}

\section{Introduction}

The application of Simon's algorithm \cite{Simon} in post-quantum cryptanalysis of block ciphers has been firstly demonstrated by  H. Kuwakado and M. Morii \cite{Hidenori} in 2010, where a quantum distinguisher for the 3-round Feistel network \cite{Feistel} has been constructed. Since then, many works have been applying Simon's algorithm in order to re-analyse the security of block ciphers and various modes on top of them in the so-called post-quantum setting. In this context, Simon's algorithm has been usually combined with Grover's \cite{Grover} algorithm, an approach introduced by G. Leander and A. May in \cite{Leander}, and it has been shown that many symmetric cryptographic schemes and general block cipher constructions provide much less security than expected. Such examples are the Even-Mansour \cite{EM-IEEE} scheme, the AEZ~\cite{AEZ} and LED ciphers~\cite{LED}, the FX construction~\cite{OfflineS,Leander}, a number of authenticated encryption schemes~\cite{QLA,Kaplan,AESCOPA}, tweakable enciphering schemes~\cite{Palash}, to name a few. In addition, many recent works such as~\cite{Cid,Palash,SHGFN3,Zhon,ZhouBV} (and references therein), deal with post-quantum cryptanalysis on reduced-round versions of Generalized Feistel schemes by applying Simon's algorithm or the Simon-Grover combination \cite{Leander} .

The main idea behind an attack based on the Simon's algorithm is to construct a periodic function $f$ using the elements of the underlying cryptographic scheme, such that the values of $f$ are available to the adversary via an oracle (as a black-box). The security model in which the adversary has access only to classical encryption (from which the values of $f$ are obtained) or decryption queries to the oracle, is known as the so-called \textbf{Q1} attack model. In the \textbf{Q2} model, the adversary has the quantum access to the encryption oracle (i.e. superposition access).
Usually, the function $f$ is constructed in such a way that its period is either the secret key (as in the case for the Even-Mansour scheme), or some other unknown value (depending on the underlying cipher), in which case $f$ is used as a quantum distinguisher due to its periodicity property. Note that the periodicity of  a random permutation is not expected with very high probability. If $f$ has only one non-trivial period, then Simon's algorithm finds the period of $f$ in quantum polynomial time $\mathcal{O}(n)$ ($n$ is the dimension of the domain of $f$), which is significantly less than the classical computational complexity $\mathcal{O}(2^{n/2})$ that would be required. 
%
%
Whenever the construction of $f$ gives rise to unwanted periods (under the assumption that there are not many periods), one can apply the results by M. Kaplan et al.~\cite{Kaplan} in order to recover the wanted period. 

\subsection*{Contributions}

The main contributions of this work can be summarized as follows:\\\\
\textbf{(1):} We apply Simon's algorithm under the \textbf{Q2} attack model to pseudorandom functions built on the Farfalle construction and their modes~\cite{Farfale}: a session-supporting mode for authenticated encryption (Farfalle-SAE), a synthetic initial value AE mode (Farfalle-SIV) and a wide block cipher mode (Farfalle-WBC). In this part, two constructions of periodic functions involving different number of blocks have been provided (that is \textbf{Constructions 1} and \textbf{2}).\\\\
More precisely, the main weakness which admits the application of Simon's algorithm to these schemes is the structure of an internal sum which is a periodic function with respect to the given input blocks. Based on this observation, we firstly show that one can provide different constructions of a periodic function with a period which contains a secret key (\Cref{sec:Farfale}). More precisely, we firstly show that one can consider a message of two blocks only (both blocks are equal and represent a variable), and construct a periodic function via \textbf{Construction 1} (\Cref{sec:constructions}). Then, this construction can be easily extended by any number of blocks, where one fixes the two blocks which are equal and represent a variable, and the rest of the blocks are constant. Another approach is to consider a message with more than two blocks, such that it has multiple pairs of equal blocks and represent different variables (eventually a constant can be added to these pairs). These two construction methods are given by \textbf{Construction 2} (\Cref{sec:constructions}). In general, the design of Farfalle suggests that one can utilize lightweight permutations internally (the $roll_c$ permutation, cf. \Cref{alg:Farfale}), and these are taking a secret value as an input. In the special case when the permutation $roll_c$ is a linear mapping (such as in Kravatte, cf.~\cite[Section 7]{Farfale}), we show (\Cref{sec:Farfaleperiod}) that one can extract the secret key under certain reasonable assumptions, which are essentially satisfied when a sum of defining matrices of different powers of $roll_c$ is an invertible mapping (or if it has a larger rank, then it may significantly reduce the key space). \\\\
\textbf{(2):} Furthermore, in \Cref{sec:Farfalemodes} we show that \textbf{Constructions 1} and \textbf{2} can be used to mount forgery attacks on the Farfalle modes SAE, SIV and WBC.\\\\
More precisely, for the Farfalle-SAE mode we demonstrate that a forgery can be done in the case when the length of the plaintext is zero, in which case one only manipulates with metadata/associated data blocks. Regarding the Farfalle-SIV mode, the forgery is possible by manipulating both metadata and/or plaintext. And finally, as the Farfalle-WBC (similarly goes for Farfalle-WBC-AE) scheme is a 4-round Feistel scheme, then a distinguishing attack can be mounted based on the fact that the input branches may contain at least two blocks (whose length is as the size of the permutation $p_c$ in Farfalle). Here, we assume that the PRFs in Farfalle-WBC are taken to be the Farfalle pseudorandom functions.\\\\
\textbf{(3):} And finally (in \Cref{sec:period}), we show that a round key involved in a period (that one obtains by applying Simon's algorithm) of a given reduced-round version of a Key-Alternating General Feistel Networks (GFN) can be efficiently extracted. In \Cref{sec:anfapproach} and \Cref{sec:lagrange} we provide two methods for round key extractions (using vector space and finite field representations respectively), which are based on different functional interpolation formulas. Moreover, in \Cref{sec:lagrange2} we provide an improvement of the round key-extraction method specifically based on the vector space interpolation, by imposing certain trade-offs regarding different parameters, such as on-line quantum query, data complexities and number of used qubits. Best to our knowledge, the extraction of a secret round key from obtained period in GFNs has not been addressed so far.\\\\
In this part, assuming that inner function $F_k$ of a given GFN is defined as $F_k(x)=F(x\oplus k)$ ($x$ is an input block, $k$ is a round key, $F$ is a publicly known function), we firstly show that for the vector space representation of $F$, one can utilize an interpolation formula which requires the knowledge of output values given at certain specific inputs (\Cref{sec:anfapproach}). Then, in context of the finite filed representation of $F$, we show that the Lagrange's interpolation formula (e.g., see \cite[Subsection 2.1.7.3]{Handbook}) can also be used in order to extract the secret round key $k$ (\Cref{sec:lagrange}). Overall, both interpolation methods require multiple application of the attack (either Simon's algorithm only, or Simon-Grover combination) by which one obtains different periods (which are used for the interpolation process). The number of this applications strongly depends on the polynomial/algebraic degree of $F$ (depending on the interpolation method), which is especially in the vector space representation usually very small in most of the existing schemes, and thus makes the extraction quite feasible. At this point, we note that the Simon-Grover combination is an algorithm which has been firstly introduced by G. Leander and A. May \cite{Leander}, and since then it has been applied to various GFNs for certain round-reduced versions which do not admit the application of Simon's algorithm only.

\subsection*{Outline}
The article is organized as follows. In \Cref{sec:pre} we provide an overview of Simon's algorithm along with some basic notation. In \Cref{sec:constructions} we demonstrate the constructions of periodic functions to Farfalle: \textbf{Constructions 1} and \textbf{2}. Then, in \Cref{sec:Farfaleperiod} we consider the extraction of a secret key from different Farfalle periods. The application of \textbf{Constructions 1} and \textbf{2} is further utilized in Section \Cref{sec:Farfalemodes}, where forgery attacks are given for Farfalle SAE and SIV, as well as a construction of a quantum distinguisher for the WBC mode. The extraction of a secret round key from different periods obtained from GFNs is shown in \Cref{sec:period}, along with the improvement method for the vector space representation of an inner function. We give our concluding remarks in \Cref{sec:OP}.

\section{Preliminaries}\label{sec:pre}

The vector space $\mathbb{F}_2^n$ is the space of all $n$-tuples $ {x}=(x_1,\ldots,x_{n})$, where $x_i \in \mathbb{F}_2$. The all-zero vector is
denoted by $\textbf{0}_n$.
%
A quantum register is a collection of $n$ qubits (the classical basis states $|0\rangle$ and $|1\rangle$), and formally we denote it as $|x\rangle=|c_1\rangle\otimes \ldots \otimes |c_n\rangle$, where $c_i\in \mathbb{F}_2$ (and thus $x\in \mathbb{F}^n_2$). 
%
An operator $U_f$, which implements a function $f:\mathbb{F}_2^n\rightarrow \mathbb{F}_2^{\tau}$ quantumly, uses the all-zero register $|\textbf{0}_{\tau}\rangle$ with auxiliary qubits as $U_f:|x\rangle|\textbf{0}_{\tau}\rangle\rightarrow |x\rangle|\textbf{0}_{\tau}\oplus f(x)\rangle=|x\rangle|f(x)\rangle.$
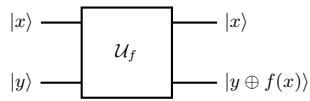
\begin{figure}[H]
\centering
\input{QuantumGate.tex}
\end{figure}
Throughout the paper, when $f$ is a function which uses a secret value (e.g. a secret key), then $\mathcal{U}_f$ will denote a quantum oracle (as unitary operator) that provides the values of $f$. 

In relation to Simon's quantum algorithm described later on, we will use the Hadamard transform $H^{\otimes n}$, $n\geq 1$ (also known as the Sylvester-Hadamard matrix), which is defined recursively as
\begin{eqnarray*}\label{HM}
 H^{\otimes 1}=2^{-1/2}\left(
                           \begin{array}{cc}
                             1 & 1 \\
                             1 & -1 \\
                           \end{array}
                         \right);\hskip 0.4cm H^{\otimes n}=2^{-1/2}\left(
      \begin{array}{cc}
        H^{\otimes (n-1)} & H^{\otimes (n-1)} \\
        H^{\otimes (n-1)} & -H^{\otimes (n-1)} \\
      \end{array}
    \right).
\end{eqnarray*}
Throughout the article, by $|\textbf{0}\rangle$ we will denote the all-zero quantum register whose size will be clear from the context. 

\subsection{On Simon's algorithm}\label{sec:Simon}

Suppose that a Boolean (vectorial) function $f:\mathbb{F}^n_2\rightarrow \mathbb{F}^\tau_2$ ($\tau\geq 1$) has a unique (secret) period $s$, that is 
$$f(x)=f(y)\;\Leftrightarrow\;x\oplus y\in\{\textbf{0}_n,s\}.$$
This problem is solved efficiently by Simon's algorithm \cite{Simon}, which extracts $s$ in $\mathcal{O}(n)$ quantum oracle queries. Classically,  solving this problem requires exponential complexity $\mathcal{O}(2^{n/2})$. In applications of this algorithm in post-quantum cryptanalysis of block ciphers (and its modes), one usually works in an environment in which a given function is not expected to have more than one period due to its complexity, or even if it has more periods, then some other techniques are applied to extract a desired period. For instance, Theorems 1 and 2 given by M. Kaplan et al. in \cite{Kaplan} show that one can still extract (with high probability) a desired shift just by performing more queries. On the other hand, if there exist other periods with probability $>1/2$, then one can apply a classical distinguishing attack based on higher order differentials with probability $>1/2$ (cf. \cite{Kaplan}). Thus, in our work we will assume that an observed function does not have many periods in general, while on the other hand, it is known that a random (vectorial) function has periods with negligible probability.

As our work utilizes Simon's algorithm as a main tool, we recall its computation steps below.\\\
\textbf{Simon's algorithm \cite{Simon}:}
\begin{enumerate}[1)]
\item Prepare the state $2^{-n/2}\sum_{x\in \mathbb{F}^n_2}|x\rangle|\textbf{0}\rangle$, where the second all-zero register is of size $\tau$ ($x\in \mathbb{F}^n_2$).
\item Apply the operator $\mathcal{U}_f$ which implements the function $f:\mathbb{F}^n_2\rightarrow \mathbb{F}^{\tau}_2$ in order to obtain the state
$2^{-n/2}\sum_{x\in \mathbb{F}^n_2}|x\rangle|f(x)\rangle.$
\item Measure the second register (the one with values of $f$), the previous state is collapsed to $|\Omega_a|^{-1/2}\sum_{x\in \Omega_a}|x\rangle$, where $\Omega_a=\{x\in \mathbb{F}^n_2:f(x)=a\}$, for some $a\in Im(f)$.
\item Apply the Hadamard transform $H^{\otimes n}$ to the state $|\Omega_a|^{-1/2}\sum_{x\in \Omega_a}|x\rangle$ in order to obtain $|\varphi\rangle=|\Omega_a|^{-1/2}2^{-n/2}\sum_{y\in \mathbb{F}^n_2}\sum_{x\in \Omega_a}(-1)^{x\cdot y}|y\rangle.$
\item Measure the state $|\varphi\rangle$:
\begin{enumerate}[1)]
\item If $f$ does not have any period, the output of the measurement are random values $y\in \mathbb{F}^n_2.$
\item If $f$ has a period $s$, the output of measurement are vectors $y$ which are strictly orthogonal to $s$, since the amplitudes of $y$ are given by

$2^{-(n+1)/2}\sum_{x\in \Omega_a}(-1)^{x\cdot y}=2^{-(n+1)/2}[(-1)^{x'\cdot y}+(-1)^{(x'\oplus s)\cdot y}],$ where the term $2^{-(n+1)/2}$ comes from the assumption that $|\Omega_a|=2$ for any $a\in \mathbb{F}^n_2.$
\end{enumerate}
\item If $f$ has a period, repeat the previous steps until one collects $n-1$ linearly independent vectors $y_i$. Then, solve the homogeneous system of equations $y_i\cdot s=0$ (for collected values $y_i$) in order to extract the unique period $s$.
\end{enumerate}
\begin{remark}\label{rem:systemCompl}
The last step above has complexity $\mathcal{O}(n^3)$, which stands for the Gaussian elimination procedure (used for solving linear systems). For complexity estimates throughout the paper, it will be neglected. Also, by Simon's function we will call a periodic function which is constructed for an observed scheme, whether it has one or more periods.  Without explicitly stating it, whenever we construct a periodic function, it will be clear that one can apply Simon's algorithm in order to obtain its period(s) (where we assume that observed scheme does not have many periods, as discussed earlier).
\end{remark}
%
A common technique to construct a Simon's function $f$ (used in many works) is to concatenate two suitable functions as shown below.
\begin{proposition}\label{pr:con}
  Let the function $f:\mathbb{F}_2\times \mathbb{F}^n_2\rightarrow \mathbb{F}^n_2$ be defined as
  \begin{eqnarray}\label{eq:f12}
  f(b,x)= \left\{\begin{array}{cc}
                   g(x), & b=0 \\
                   h(x), & b=1
                 \end{array}
  \right.,
  \end{eqnarray}
where $g,h:\mathbb{F}^n_2\rightarrow \mathbb{F}^n_2.$  Then, if $h(x\oplus s)=g(x)$ holds for all  $x\in \mathbb{F}^n_2$, then $f$ has period $(1,s)\in \mathbb{F}_2\times \mathbb{F}^n_2.$
\end{proposition}
\begin{remark}\label{rem:uniqueS}
Note that if $g$ and $h$ in \Cref{eq:f12} are permutations, then using the same arguments as in the proof of Lemma 1 in \cite{Hidenori}, one can show that $(1,s)$ is an unique period of $f$.
\end{remark}
In what follows we recall two examples of constructions of $f$ provided in previous works, which are special cases of Proposition \ref{pr:con}.
\begin{example}\label{ex:e1}
In \cite{Hidenori} authors analysed the distinguishability of 3-round Feistel cipher, and in Section III-A defined the function $f:\mathbb{F}_2\times \mathbb{F}^{n}_2\rightarrow \mathbb{F}^n_2$ by
\begin{eqnarray}\label{eq:Feistel3}
f(b,x)=\left\{\begin{array}{cc}
                  F_2(x\oplus F_1(\alpha))\oplus (\alpha\oplus \beta), & b=0 \\
                 F_2(x\oplus F_1(\beta))\oplus (\alpha\oplus \beta), & b=1
                \end{array}
\right.,
\end{eqnarray}
where $\alpha,\beta\in \mathbb{F}^n_2$ are different fixed vectors, and $F_i$ denotes an inner function of the 3-round Feistel cipher which depends on the round key $k_i$, i.e. $F_i=F_{k_i}$, $i=1,2$. Denoting by $g(x)=f(0,x)$ and $h(x)=f(1,x)$, it is easily verified that $h(x\oplus (F_1(\alpha) \oplus F_1(\beta)))=g(x)$, and thus by Proposition \ref{pr:con}, $f$ has a linear structure $(1,s)=(1,F_1(\alpha) \oplus F_1(\beta))\in \mathbb{F}_2\times \mathbb{F}^{n}_2$.
\end{example}
\begin{remark}
Note that there exist several other works who use the similar ideas to construct the Simon's function $f$ as above (for instance, \cite[Section 3.2]{Dong}, \cite[Section 5.1]{Kaplan}, \cite{LED}, etc.).
\end{remark}
%
On the other hand, there are constructions of the function $f$ which are not based on concatenation method. For instance, in \cite{EM-IEEE} the Even-Mansour cipher has been broken (in the setting which uses two keys), which is given by $E_{k_1,k_2}(x)=P(x\oplus k_1)\oplus k_2$, by constructing the following function $f:\mathbb{F}^{n}_2\rightarrow \mathbb{F}^n_2$ by
$$f(x)=E_{k_1,k_2}(x)\oplus P(x)=P(x\oplus k_1)\oplus P(x)\oplus k_2.$$
One can easily verify that $f(x\oplus k_1)=f(x)$ holds for all $x\in \mathbb{F}^{n}_2$ (i.e., we have that $s=k_1$). Further analysis and recovering of the keys $k_1$ and $k_2$ one can find in \cite[Section 3.2]{Kaplan}. 

Another example is LRW construction \cite{Ronald} analysed in \cite[Section 3.2]{Kaplan}, which is defined as $\tilde{E}_{t,k}(x)=E_k(x\oplus d(t))\oplus d(t)$, where $d$ is a universal hash function. Constructing the function $f:\mathbb{F}^{n}_2\rightarrow \mathbb{F}^n_2$ by
$$f(x)=E_k(x\oplus d(t_0))\oplus d(t_0)\oplus E_k(x\oplus d(t_1))\oplus d(t_1),$$
where $t_0, t_1$ are two arbitrary distinct tweaks, we have that $f$ is periodic in $s=d(t_0)\oplus d(t_1)$. 
\begin{remark}
In \Cref{sec:Farfale}, we will be mainly utilizing the construction approach similar to the one used for the LRW scheme, as the pseudorandom function Farfalle internally admits the same weakness. Essentially, we see that the LRW construction admits the swapping of (secret) terms present with the variable $x$ (that is $d(t_0)$ and $d(t_1)$). Note that the same construction is used for other schemes (e.g. PMAC, OCB, etc.), see for instance \cite{Kaplan}.
\end{remark}

\section{Applying Simon's algorithm to Farfalle}\label{sec:Farfale}

\begin{figure}[h!]
  \centering
  \scalebox{0.9}{
\input{farfalle.tikz}}
\caption[The Farfalle construction]{The Farfalle construction.\protect\footnotemark}
\label{fig:Farfale}
\end{figure}
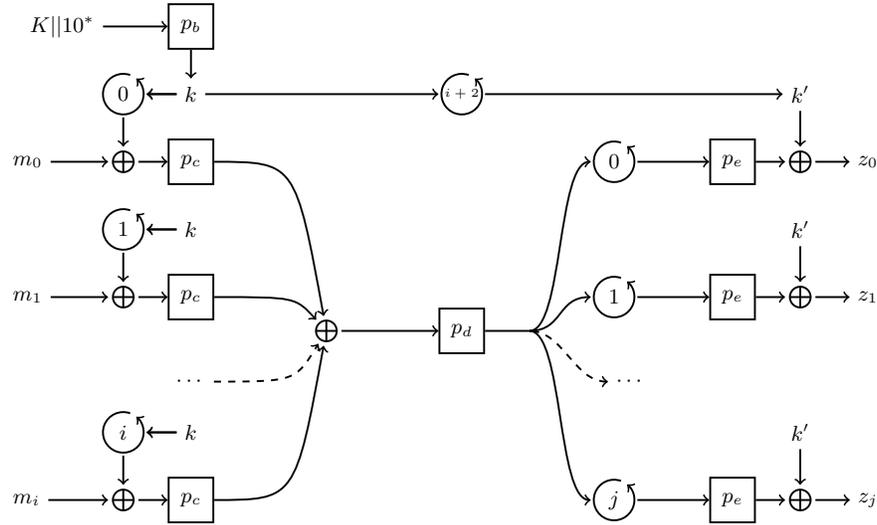

Farfalle is a permutation-based construction for building a pseudorandom function (PRF) which has been introduced by G. Bertoni et al.~\cite{Farfale} in 2017. The construction of Farfalle is given in \Cref{fig:Farfale}, while the precise description of the scheme is given in \Cref{alg:Farfale}.

\begin{algorithm}
  \renewcommand{\algorithmicrequire}{\textbf{Parameters:}}
  \caption{The Farfalle construction. \cite{Farfale}} \label{alg:Farfale}
  \begin{algorithmic}
    \Require $b$-bit permutations $p_b, p_c, p_d, p_e$ and rolling functions $roll_c$, $roll_e$ ($F$ below denotes the output of Farfalle)
    \renewcommand{\algorithmicrequire}{\textbf{Input:}}
    \renewcommand{\algorithmicensure}{\textbf{Output:}}
    \Require
    \State key $K \in \Z_2^*, |K| \leq b-1$
    \State input string sequence $M^{(t-1)} \circ \ldots \circ M^{(0)} \in (\Z_2^*)^{+}$
    \State length $n \in \N$ and offset $q \in \N$
    \Ensure string $Z \in \Z_2^n$
    \State $K' \gets \textsf{pad} 10^*(K)$
    \State $k \gets p_b(K')$ \algorithmiccomment{mask derivation}
    \State $x \gets 0^b$, $I \gets 0$
    \For{$0 \ldots t-1$}
    \State $M \gets \textsf{pad} 10^*(M^{(j)})$
    \State Split $M$ in $b$-bit blocks $m_{I}$ to $m_{I + \mu -1}$
    \State $x \gets x + \sum_{i = 1}^{I + \mu - 1} p_c(m_i + roll_c^i(k) )$
    \State $I \gets I + \mu + 1$ \algorithmiccomment{skip the blank index}
    \EndFor
    \State $k' \gets roll_c^I(k)$, $y \gets p_d(x)$
    \While{all the requested $n$ bits are not yet produced}
    \State produce $b$-bit blocks as $z_j = p_e(roll_e^j(y)) + k'$
    \EndWhile
    \State $Z \gets n$ successive bits from $z_0\Vert z_1 \Vert z_2 \ldots$ starting from bit with index $q$
    \State \Return{$Z = 0^n + F(M^{(t-1)} \circ \ldots \circ M^{(0)}) \lll q$} 
  \end{algorithmic}
\end{algorithm}
In general, the construction of Farfalle employs (lightweight) $b$-bit permutations $p_b, p_c, p_d$ and $p_e$. In addition, the rolling functions $roll^i_c$ and $roll^i_e$ (also permutations) are used for compression and expansion, respectively, where by construction these two functions should be chosen such that an adversary not knowing the value $k$ shall not be able to predict the mask values $roll^i_c(k)$ for any $i$ (in a reasonable range), nor the difference between any pair of mask values $roll^i_c(k)\oplus roll^j_c(k)$ for any $i\neq j$.  

We recall that two instances of the Farfalle construction are known as Kravatte, which was presented in \cite{Farfale} and which is based on Keccak-$p[1600, n_r]$ permutations \cite{Keccak, fips202}, and Xoofff, which is  \cite{Xoofff} based on Xoodoo \cite{Xoodoo} permutation with specific rolling functions (cf. \cite[Definition 2]{Xoofff}). In what follows, we consider different cases for which one can construct a periodic function by utilizing outputs of Farfalle (to which Simon's algorithm applies), which here will be output blocks $z_j$. 
\begin{remark}
In the context of the Kravatte and Xoofff instances, we note that their authors do not claim quantum resistance against an attacker who can make quantum superposition queries (the \textbf{Q2} attack model), i.e. the presented results in subsequent sections do not contradict their security claims.
\end{remark}
\begin{remark}
For convenience, in our constructions we will assume that all message blocks $m_i$ are directly corresponding to the $roll^i_c$ functions in Farfalle, i.e. there are no \emph{blank indices} which are not contributing  to accumulators (cf. \cite[Figure 2]{Farfale}). Note that this assumption does not affect our construction methods in general, since the messages could always be chosen suitably so that they yield periodicity of output blocks $z_j$.
\end{remark}

\footnotetext{Figure generated with modified TikZ script from \cite{tikz-iacr}}

\subsection{Constructing periodic functions}\label{sec:constructions}

We present the following constructions:\\\\
\textbf{Construction 1:} $\boldsymbol{a)}$ Let $M$ be a message which contains only two $b$-bit blocks $m_i$, namely $M=m_0||m_1$. Following the notation of the Farfalle algorithm, we have that the internal value (denoted by $Y$) is given as
$$Y=p_d[p_c(m_0\oplus roll^0_c(k))\oplus p_c(m_1\oplus roll^1_c(k))].$$
Now, if we set the blocks $m_0$ and $m_1$ to be equal and represent the same variable, that is $m_0=m_1=m\in \mathbb{F}^b_2$ (note that $p_c:\mathbb{F}^b_2\rightarrow \mathbb{F}^b_2$), we have that any output block $z_j$ ($j\geq 0$), is given as
$$z_j(m)=p_e(roll^j_e(Y))\oplus k'=p_e(roll^j_e(p_d[p_c(m\oplus roll^0_c(k))\oplus p_c(m\oplus roll^1_c(k))]))\oplus k',$$
where $k'$ is a constant since it depends on the constant $k$ and the number of input blocks (which is related to $i$). We note that a value $j$ can be chosen arbitrarily, which falls into the range of the first $q$ bits in the last step of the Farfalle algorithm.

Now, we observe that the function $z_j=z_j(m)$ is periodic with the period $s=roll^0_c(k)\oplus roll^1_c(k)$, i.e. it holds that:
$$z_j(m\oplus roll^0_c(k)\oplus roll^1_c(k))=z_j(m),\;\;\;\forall m\in \mathbb{F}^b_2.$$
Consequently, by applying Simon's algorithm to $z_j(m)$ one is able to extract the value $s$ in quantum polynomial time $\mathcal{O}(b)$. We note that $s$ is not expected to be equal to $\textbf{0}_b$, due to the design criterions for the $roll_c$ permutation.\\\\
$\boldsymbol{b)}$ Similarly, we consider the message $M=(m\oplus \alpha)||(m\oplus \beta)$ with $m$ being a variable and arbitrary fixed values $\alpha,\beta\in \mathbb{F}^b_2$, in which case any output function $z_j(m)$ is given as
$$z_j(m)=p_e(roll^j_e(p_d[p_c(m\oplus  \alpha\oplus roll^0_c(k))\oplus p_c(m\oplus \beta\oplus roll^1_c(k))]))\oplus k'.$$
Clearly, $z_j(m)$ has the period $\alpha\oplus \beta\oplus roll^0_c(k)\oplus roll^1_c(k),$ which can be extracted by Simon's algorithm in polynomial time $\mathcal{O}(b)$.\\\\
%
%
%
%
\textbf{Construction 2:} The previous construction can be extended to the case when the message $M$ contains more blocks. Consider the following cases:\\\\
$\textbf{i)}$ For instance, one may take two blocks to be equal and represent a variable, while the remaining blocks are all arbitrary constants. More precisely, such a message is given by
$$M=m_0||m_1||\alpha_1||\ldots||\alpha_t=m||m||\alpha_1||\ldots||\alpha_t,\;\;m\in \mathbb{F}^b_2,$$
where $\alpha_u \in \mathbb{F}^b_2$ are arbitrary constants. In this case, the internal value $Y$ is given as
$$Y=p_d(\boldsymbol{p_c(m\oplus roll^0_c(k))}\oplus \boldsymbol{p_c(m \oplus roll^1_c(k))}\oplus \bigoplus^{t}_{i=0} p_c(\alpha_i\oplus roll^i_c(k))),$$
which again implies the periodicity of the output function (block) $z_j(m)$ (for any $j$), with the period $roll^0_c(k)\oplus roll^1_c(k).$ This holds due to the fact that the sum $\bigoplus^{t}_{i=0} p_c(\alpha_i\oplus roll^i_c(k))$ is constant and does not affect the periodicity of $z_j(m).$\\\\
$\textbf{ii)}$ Another approach here is to consider a message given as
$$M=(m_0\oplus \alpha_0)||(m_0\oplus \alpha_0)||(m_1\oplus \alpha_1)||(m_1\oplus \alpha_1),$$
where $m_r\in \mathbb{F}^b_2$ are variables and  $\alpha_r \in \mathbb{F}^b_2$ ($r=0,1$) are arbitrary constants. In this case, due to the periodicity of values
$$Y_r=p_c(m_r\oplus \alpha_r\oplus roll^{0+2r}_c(k))\oplus p_c(m_r\oplus \alpha_r\oplus roll^{1+2r}_c(k)),\;\;r=0,1,$$
we have that the output function 
$$z_j(m_0,m_1)=p_e(roll^j_e(Y_0\oplus Y_1))\oplus k'$$ 
is periodic, with periods $(roll^0_c(k)\oplus roll^1_c(k),\textbf{0}_b)$ and $(\textbf{0}_b,roll^2_c(k)\oplus roll^3_c(k))$, i.e. $z_j(m_0,m_1)$ is periodic in both arguments.
\begin{remark}
Clearly, the construction above can be generalized by taking more than two distinct variables which will be present two times in the message (in $M$ given above we have two times $m_0$ and two times $m_1$). 
\end{remark}
The demonstrated constructions show that:
\begin{itemize}
\item The periodicity of functions $z_j(m)$ in all previous constructions does not depend on the choice of rolling functions $roll_c$ and $roll_e$, nor on the choice of permutations $p_b, p_c, p_d$ and $p_e$.
\item One can choose a suitable message block $M=\alpha||m||\beta||m||\gamma$ ($\alpha,\beta,\gamma$ fixed constants of certain lengths) which contains the variable block $m$ placed such that it corresponds to two different indices, say $i$ and $j$ ($i<j$), in which case the period of any output function $z_t(m)$ ($t\geq 0$) is given as $s=roll^i_c(k)\oplus roll^j_c(k)$. This plays an important role in the process of the key extraction (as shown in \Cref{sec:Farfaleperiod}).
\end{itemize}

\subsection{Extracting the secret key $K$ in Farfalle}\label{sec:Farfaleperiod}

As discussed in \cite{Farfale},  the permutation $p_c$ can be instantiated by some lightweight functions (say, with degree equal to $2$). In this context, its resistance to higher-order differential attacks has been discussed  in \cite[Section 8.1]{Farfale}, where it has been noted that whenever an adversary wants to construct four message blocks $m_{i_1},m_{i_2},m_{i_3},m_{i_4}$ such that $m_{i_1}\oplus m_{i_2}\oplus m_{i_3}\oplus m_{i_4}=\textbf{0}$, then the difficulty of it relies on the property that the right-hand side of the equality
$$m_{i_1}\oplus m_{i_2}\oplus m_{i_3}\oplus m_{i_4}=roll^{i_1}(k)\oplus roll^{i_2}(k)\oplus roll^{i_3}(k)\oplus roll^{i_4}(k)$$
should not result in a linear mapping whose defining matrix has a low degree. 

More precisely, considering that $roll$ is a linear function defined as  $roll(k)=\mathcal{M}\times k$ (for some invertible binary matrix $\mathcal{M}$), the right-hand side can be written as $(\mathcal{M}^{i_1}\oplus \mathcal{M}^{i_2}\oplus \mathcal{M}^{i_3}\oplus \mathcal{M}^{i_4})\times k$, and thus guessing  its value becomes more difficult if the matrix $\mathcal{M}^{i_1}\oplus \mathcal{M}^{i_2}\oplus \mathcal{M}^{i_3}\oplus \mathcal{M}^{i_4}$ has full rank (or higher rank in general), for any four pairwise indices $i_1,i_2,i_3,i_4$ in some reasonable range which limits the maximum number of blocks in Farfalle. Consequently, this gives a design requirement for the rolling function $roll_c$ in general.

Regarding our \textbf{Constructions 1} and \textbf{2}, we note that the extraction of the secret value $k$ is in a trade-off with the previous requirement. Namely, by applying the given constructions, one can deduce the periods of the form 
$$s_{i,j}=roll^i_c(k)\oplus roll^j_c(k),\;\;\;i\neq j,$$ which corresponds to pairwise distinct variable blocks with indices $i$ and $j$. Assuming that $roll_c$ is an invertible linear function defined as $roll_c(k)=\mathcal{M}\times k$, then one can combine different periods $s_{i_t,j_t}$ in order to derive the sum
\begin{eqnarray}\label{eq:systemK}
s_{i_1,j_1}\oplus \ldots \oplus s_{i_p,j_p}=\bigoplus^p_{t=1}(\mathcal{M}^{i_t}\oplus \mathcal{M}^{j_t})\times k,\;\;\;i_t\neq j_t.
\end{eqnarray}
For instance, one may derive the period $s_{0,1}=roll^0_c(k)\oplus roll^1_c(k)$, or $s_{0,2}=roll^0_c(k)\oplus roll^2_c(k)$, etc. Recall that any period $s_{i_t,j_t}$ is obtained in quantum polynomial time $\mathcal{O}(b)$ from Farfalle, by applying Simon's algorithm. Consequently, we have that $k$ is a solution of the linear system (\Cref{eq:systemK}), and thus a uniqe $k$ exists if $\bigoplus^p_{t=1}(\mathcal{M}^{i_t}\oplus \mathcal{M}^{j_t})$ is an invertible matrix. Clearly, the space of possible solutions of (\Cref{eq:systemK}) may be reduced depending on the rank of the matrix $\bigoplus^p_{t=1}(\mathcal{M}^{i_t}\oplus \mathcal{M}^{j_t})$. As the function $p_b$ is a permutation in Farfalle, this discussion is formalized in the following result.
\begin{proposition}\label{prop:systemK}
Let $s_{i_t,j_t}=roll^{i_t}_c(k)\oplus roll^{j_t}_c(k)$ ($1\leq i_t< j_t\leq p$) be periods obtained from Farfalle by applying Simon's algorithm. 
Suppose that the $roll_c(k)$ is a linear function defined as $roll_c(k)=\mathcal{M}\times k.$ Then, the secret value $k$ is solution of the linear system of equations (\Cref{eq:systemK}). Consequently, the secret key $K$ in Farfalle can be determined from $K||10^*=p^{-1}_b(k)$.
\end{proposition}
\begin{remark}
Note that the complexity of obtaining $p$ periods $s_{i_t,j_t}$ is $\mathcal{O}(pb)$ (cf. Remark \cref{rem:systemCompl}), which clearly depends  on the requirement that $\bigoplus^p_{t=1}(\mathcal{M}^{i_t}\oplus \mathcal{M}^{j_t})$ has a full rank. In this context, one may further consider the work \cite{Muller},  which analyses the invertibility of a sum of two nonsingular matrices.
\end{remark}

\subsection{Attacking authenticated encryption modes based on Farfalle}\label{sec:Farfalemodes}

Based on the previously presented constructions we now consider attacks on certain authenticated encryption modes based on Farfalle, namely Farfalle-SAE and Farfalle-SIV. These modes apply a pseudorandom function (PRF) $F$ that is instantiated as the Farfalle PRF. Without mentioning explicitly, the extraction of the secret key $K$ will be possible whenever one meets the requirements of \Cref{prop:systemK}.\\\\
\textbf{Farfalle-SAE} is a session-supporting authenticated encryption scheme, where the initialization and wrapping steps are defined as in Algorithm \ref{alg:FarfaleSAE}.

\begin{algorithm}
  \renewcommand{\algorithmicrequire}{\textbf{Parameters:}}
  \caption{Farfalle-SAE[$F, t, \ell$] \cite{Farfale}} \label{alg:FarfaleSAE}
  \begin{algorithmic}
    \Require PRF $F$, tag length $t \in \N$, and alignment unit length $\ell \in \N$

    \State \textbf{Initialization} takes $K \in \Z_2^*$, nonce $N \in \Z_2^*$ and returns tag $T \in \Z_2^t$
    \State offset = $\ell \lceil \frac{t}{\ell}\rceil$: smallest multiple of $\ell$ not smaller than $t$
    \State $history \gets N$
    \State $T \gets 0^t + F_K(history)$
    \State \Return $T$

    \State \textbf{Wrap} takes metadata $A \in \Z_2^*$, plaintext $P \in \Z_2^*$, returns ciphertext $C \in \Z_2^{|P|}$ and tag $T \in \Z_2^t$
    \State $C \gets P + F_K(history) \ll \textrm{offset}$
    \If{$|A| > 0$ OR $|P| = 0$}
    \State  $history \gets A \Vert 0 \circ history$
    \EndIf
    \If{$|P| > 0$}
    \State  $history \gets C \Vert 1 \circ history$
    \EndIf
    \State $T \gets 0^t + F_K(history)$
    \State \Return $C, T$
  \end{algorithmic}
\end{algorithm}
Now, following the wrapping procedure of Farfalle-SAE, let us assume that $|A|>0$ and $|P|=0$, i.e. we are considering the case when there is no plaintext. Still, in this case we assume that the nonce $N$ changes with every new metadata $A$. In the case when $|A|>0$ and $|P|=0$, we have that $history$ updates as $history\leftarrow A||0\circ history=A||0\circ F_K(N)$, and thus let us assume that the metadata $A$ is given by 
$$A=a_0||a_1=a||a,\;\;\;a\in \mathbb{F}^b_2,$$
where $b$ is length of the $p_c$ permutation used in Farfalle. In the last step, we have that the tag $T$ is updated as 
$$T\leftarrow F_K(A||0\circ F_K(N)),$$
which is clearly periodic in $s=roll^0_c(k)\oplus roll^1_c(k)$ (cf. \textbf{Construction 1}). This is due to the fact that the period $s$ does not depend on the "old" value of $history=F_K(N)$, i.e. $T(a)=T(a\oplus s)$ holds for any $a$ even when the nonce value changes with respect to $A$. Note that similar scenario is present in other modes as well, see for instance \cite{Kaplan}.\\\\
\textbf{Mounting a forgery attack (the case $|A|>0$ and $|P|=0$):}
\begin{enumerate}[1)]
\item Query the Farfalle-SAE oracle $\mathcal{O}_{Farfalle-SAE}$ with $A=a_0||a_1=a||a$ and no plaintext sufficiently many times until a period $s=roll^0_c(k)\oplus roll^1_c(k)$ is extracted by Simon's algorithm from any block $z_j(a)$ (for any $j$) of the tag function $T(a)=F_K(A||0\circ F_K(N))$. As any $z_j(a)$ has the same period, one can consider the whole output value of $T(a).$ This stage requires $\mathcal{O}(b)$ queries, and admits the nonce value $N$ to be different with every new value of $a$.
\item For an arbitrary (fixed) $a\in \mathbb{F}^b_2$, construct a valid tag $T'$ for the metadata $A'=(a\oplus s)||(a\oplus s)$. 
\end{enumerate}
In the case of the forgery attack, it is clear that the existence of more periods (except those that are expected) simply means new/different forgeries (i.e., it does not affect the success of the attack in a negative way). This is also the case for forgery attacks presented later for the  Farfalle-SIV mode.\\\\
%
%
%
\textbf{Farfalle-SIV} is an authenticated encryption schemes which can securely encipher different plaintexts with the same key  (Algorithm \ref{alg:FarfaleSIV}). As it uses the tag computed over the message as a nonce for the encryption function, then the security narrows down to the case when two messages have the same tag.
\begin{algorithm}
  \renewcommand{\algorithmicrequire}{\textbf{Parameters:}}
  \caption{Farfalle-SIV algorithm \cite{Farfale}} \label{alg:FarfaleSIV}
  \begin{algorithmic}
    \Require a PRF $F$ and tag length $t \in \mathbb{N}$
    \State \textbf{Wrap} takes metadata $A \in \Z_2^*$ and plaintext $P \in \Z_2^*$, and returns ciphertext $C \in \Z_2^*$ and tag $t \in \Z_2^*$
    \State $T \gets 0^t + F_K(P \circ A)$
    \State $C \gets P + F_K(T \circ A)$
    \State \Return $C,T$

    \State \textbf{Unwrap} takes metadata $A \in \Z_2^*$, ciphertext $C \in \Z_2^*$ and tag $T \in \Z_2^t$ and returns plaintext $P \in \Z_2^{|C|}$ or an error 
    \State $P \gets C + F_K(T \circ A)$
    \State $T' \gets 0^t + F_K(P \circ A)$
    \If{$T' = T$}
    \State  \Return $P$
    \Else
    \State \Return error
    \EndIf
  \end{algorithmic}
\end{algorithm}
Let us consider the following attack which is targeting the tag $T$. 

As it has been shown in \textbf{Constructions 1} and \textbf{2}, one can construct different messages $P_1$ and $P_2$, as well as the corresponding metadata blocks $A_1$ and $A_2$ such that $F(P_1\circ A_1)=F(P_2\circ A_2)$. For instance, considering \textbf{Construction 1}, one may choose plaintexts $P_i$ ($i=1,2$) as functions of $m\in \mathbb{F}^b_2$ such that
\begin{eqnarray}\label{eq:PA}
\begin{array}{c}
           P_1=m||m,\;\;\;A_1=a||a, \\
           P_2=(m\oplus s')||(m\oplus s'),\;\;\;A_2=(a\oplus s'')||(a\oplus s''),
         \end{array}
\end{eqnarray}
where $m,a\in \mathbb{F}^b_2$ are considered to be variables, $s'=roll^0_c(k)\oplus  roll^1_c(k)$ and $s''=roll^2_c(k)\oplus  roll^3_c(k)$. Consequently, we have that the internal sums $S_{P_1\circ A_1}$ and $S_{P_2\circ A_2}$ are given as
$$S_{P_1\circ A_1}=p_c(m\oplus roll^0_c(k))\oplus p_c(m\oplus roll^1_c(k))\oplus p_c(a\oplus roll^2_c(k))\oplus p_c(a\oplus roll^3_c(k)),$$
and
\begin{eqnarray*}
S_{P_2\circ A_2}&=&p_c(m\oplus s'\oplus roll^0_c(k))\oplus p_c(m\oplus s'\oplus roll^1_c(k))\oplus p_c(a\oplus s'' \oplus roll^2_c(k))\\
&\oplus& p_c(a\oplus s''\oplus roll^3_c(k)).
\end{eqnarray*}
Thus, any output block $z_j(m,a)$ of $F(P_i\circ A_i)$ ($i=1,2$) is given as
$$z^{P_1\circ A_1}_j(m,a)=p_e(roll^j_e(p_d[S_{P_1\circ A_1}]))\oplus k'=p_e(roll^j_e(p_d[S_{P_2\circ A_2}]))\oplus k'=z^{P_2\circ A_2}(m,a),$$
since the vectors $s'$ and $s''$ are periods of $z^{P_1\circ A_1}_j(m,a)$, i.e. it holds that
$$z^{P_1\circ A_1}_j(m,a)=z^{P_1\circ A_1}_j(m\oplus s',a\oplus s'')=z^{P_2\circ A_2}_j(m,a).$$
Moreover, it holds that $z^{P_1\circ A_1}_j(m,a)=z^{P_1\circ A_1}_j(m\oplus s',a)$ and $z^{P_1\circ A_1}_j(m,a)=z^{P_1\circ A_1}_j(m,a\oplus s'')$, i.e. $z^{P_1\circ A_1}_j(m,a)$ is periodic in both arguments. 
\begin{remark}
It is well known that if a given function has some periods, say $s_1,\ldots,s_r$, then the same function is periodic in any value from the space spanned by $s_1,\ldots,s_r$. In context of the previous construction, the function $z^{P_1\circ A_1}_j(m,a)$ is periodic in every element of the set $\langle(s',\textbf{0}_b),(\textbf{0}_b,s'')\rangle=\{\textbf{0}_{2b},(s',\textbf{0}_b),(\textbf{0}_b,s''),$ $(s',s'')\}$, where $\textbf{0}_{2b}$ is clearly the trivial period.
\end{remark}
\textbf{On attack feasibility:} In general, the previous construction of the same tag for different inputs $P_i\circ A_i$ is possible if the attacker knows at least one of the vectors $s'$ or $s''$ (depending on whether we want to manipulate plaintext parts or metadata parts respectively). There exist two possible approaches in recovering $s'$ or $s''$:\\\\
\textbf{I)} In this approach, let us assume that  the input plaintext $P_1$ and $A_1$ have the forms given by (\Cref{eq:PA}). The attacker may firstly apply Simon's algorithm to the output function $z^{P_1\circ A_1}_j(m,a)$, which is periodic in both arguments, and then he may construct an input $P_2\circ A_2$ (as in \Cref{eq:PA}) for which the tags $T_{P_1\circ A_1}$ and $T_{P_2\circ A_2}$ are equal.\\\\
\textbf{Mounting a forgery attack ($P$ and $A$ defined as in (\Cref{eq:PA})):}
\begin{enumerate}[1)]
\item Query the Farfalle-SIV oracle $\mathcal{O}_{Farfalle-SIV}$ with inputs $P=m||m$ and $A=a||a$ sufficiently many times until the space of periods $\langle(s',\textbf{0}_b),(\textbf{0}_b,s'')\rangle$ is obtained by Simon's algorithm from any block $z_j(m,a)$ (for any $j$) of the tag function $T(m,a)=F_K(P\circ A)$. 
\item For arbitrary (fixed) blocks $m,a\in \mathbb{F}^b_2$, construct a valid tag $T'$ for the input $P'\circ A'=(m\oplus s')||(m\oplus s')||(a\oplus s'')||(a\oplus s'')$. 
\end{enumerate}
\textbf{II)} Now, if we assume that $P_1=m_0||m_1$ is any plaintext which consists of two blocks (here $m_0$ and $m_1$ are not necessarily the same), then one may assume that for all such messages the metadata $A_1$ has the form $A_1=a||a$. Then, any output block $z_j(m_0,m_1,a)$ has the period $(\textbf{0}_b,\textbf{0}_b,s'')=(\textbf{0}_b,\textbf{0}_b,roll^2_c(k)\oplus  roll^3_c(k))$, i.e. it holds that
$$z_j(m_0,m_1,a)=z_j(m_0,m_1,a\oplus s''),\;\;\forall a,m_0,m_1\in \mathbb{F}^b_2.$$
\textbf{Mounting a forgery attack (Any $P$ and $A$ is defined as in (\Cref{eq:PA})):}
\begin{enumerate}[1)]
\item Query the Farfalle-SIV oracle $\mathcal{O}_{Farfalle-SIV}$ with $A=a||a$ and arbitrary inputs $P=m_0||m_1$ sufficiently many times until the  period $s''$ is obtained by Simon's algorithm from any block $z_j(a)$ (for any $j$) of the tag function $T(a)=F_K(P\circ A)$. 
\item For arbitrary (fixed) block $a\in \mathbb{F}^b_2$, construct a valid tag $T'$ for the input $P\circ A'=m_0||m_1||(a\oplus s'')||(a\oplus s'')$. 
\end{enumerate}
\begin{remark}
Note that the versatility of the previously given forgery attacks can be achieved by placing variable blocks of $P$ and $A$ to correspond to different indices, which in turn gives different periods (as discussed at the end of \Cref{sec:constructions}).
\end{remark}
%
%
%
%
%
%
\textbf{Farfalle-WBC} is a tweakable wide block cipher (based on two PRFs) whose construction represents an instantiation of the HHFHFH mode, which was presented in \cite{HHF}. Essentially, it is a 4-round Feistel scheme which processes an arbitrary-length plaintext (\Cref{alg:WBC}). 
%
%
\begin{algorithm}
  \renewcommand{\algorithmicrequire}{\textbf{Parameters:}}
  \caption{Farfalle-WBC[$H,G, \ell$] \cite{Farfale}} \label{alg:WBC}
  \begin{algorithmic}
    \Require PRFs $H, G$ and alignment unit length $\ell \in N$

    \State \textbf{Encipher} takes $K \in \Z_2^*$, tweak $W \in \Z_2^*$, and plaintext $P \in \Z_2^*$, returns ciphertext $C \in \Z_2^{|P|}$
    \State $L \gets $  first split($|P|$) of $P$, and $R$ gets the remaining bits
    \State $R_0 \gets R_0 + H_K(L \Vert 0)$, where $R_0$ is the first $\min (b, |R|)$ bits of $R$
    \State $L \gets L + G_K(R \Vert 1 \circ W)$
    \State $R \gets R + G_K(L \Vert 0 \circ W)$
    \State $L_0 \gets L_0 + H_K(R \Vert 1)$ where $L_0$ is the first $\min(b, |L|)$ bits of $L$
    \State \Return $C = L \Vert R$
  \end{algorithmic}
\end{algorithm}
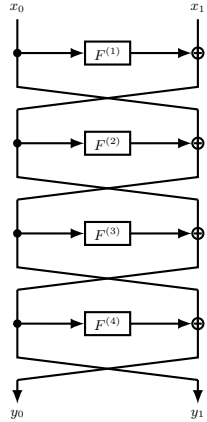
\begin{figure}[h!]
\centering
\input{Feistel1.tex} 
\caption{The 4-round Feistel network with keyed inner functions $F^{(i)}$.}
\label{fig:WBCFeistel}
\end{figure}%
In order to provide a general analysis of the security of Farfalle-WBC in terms of Simon's algorithm, for convenience we consider a 4-round Feistel scheme (\Cref{fig:WBCFeistel}) with the following setting:
\begin{itemize}
\item Inner functions $F^{(i)}$ ($i=1,\dots,4$) represent $H_K$ and $G_K$ such that $F^{(1)}(x)=H_K(x||0)$, $F^{(2)}(x)=G_K(x||W||1), F^{(3)}(x)=G_K(x||W||0), F^{(4)}(x)=H_K(x||1)$, where $W\in \mathbb{F}^*_2$ is a tweak. Note that the notation $\upsilon \circ W$ in \Cref{alg:WBC} represents the concatenation $W||\upsilon$, $\upsilon\in \{0,1\}$ (cf. the notation in Subsection 2.1 in \cite{Farfale}).
\item Inner functions $G_K$ and $H_K$ used in Farfalle-WBC algorithm are taken to be Farfalle functions.
\item An attack that we present later assumes that the size of input blocks $P_i$ ($i=1,2$) is equal to two, i.e. both branches are containing two $b$-blocks ($b$ is length of the inner permutation $p_c$ in Farfalle). It is important to note that this assumption is based on the definition of the $split[b,\ell]$ function given by \cite[Algorithm 4]{Farfale}, which admits more than one $b$-bit block per branch (for suitable underlying parameters).
\end{itemize}
With respect to the previous assumptions, we describe the Farfalle-WBC algorithm in a convenient way. Now, by observing the output value $C_2$ (left input to fourth round), which is given as
$$C_2(P_1,P_2)=P_2\oplus F^{(1)}(P_1)\oplus F^{(3)}(P_1\oplus F^{(2)}(P_2\oplus F^{(1)}(P_1))),$$  
where $P_1$ and $P_2$ represent the blocks of an input plaintext, we can construct a Simon's function $f$ as follows. Taking that $P_2=X=m||m$ ($m\in \mathbb{F}^b_2$) is a variable and $P_1=\alpha\in \mathbb{F}^{2b}_2$ is a fixed constant, we define the function $f:\mathbb{F}^b_2\rightarrow \mathbb{F}^{\tau}_2$ ($\tau$ is length of the output of $H_K$ or $G_K$ in Farfalle-WBC) as
\begin{eqnarray*}\label{eq:funkWBC}\nonumber
f(m)&=&P_2\oplus C_2(\alpha,X)=F^{(1)}(\alpha)\oplus F^{(3)}(\alpha\oplus F^{(2)}(X\oplus F^{(1)}(\alpha)))\\
&=&F^{(1)}(\alpha)\oplus F^{(3)}(\alpha\oplus F^{(2)}(m\oplus \beta_1||m\oplus \beta_2)),
\end{eqnarray*}
where $F^{(1)}(\alpha)=\beta_1||\beta_2$. We deduce the following:
\begin{enumerate}[1)]
\item We have  that the period of the internal function $$m\rightarrow F^{(2)}(m\oplus \beta_1||m\oplus \beta_2)=G_K(m\oplus \beta_1||m\oplus \beta_2||W||1)$$ is given by $s=\beta_1\oplus \beta_2\oplus roll^2(k)\oplus roll^3(k)$. Note that the blocks of $P_2$ correspond to indices $2$ and $3$ in Farfalle. This is due to the fact that every output block of the Farfalle function is periodic with the same period $s$ (visible on both \textbf{Constructions 1} and \textbf{2}), and moreover,  the tweak value $W$ (considered to be a constant) by \textbf{Construction 2-(i)} does not affect the value of the period.

\item Consequently, as other parts of the function $f$ are constant (referring to $F^{(1)}(\alpha)$ and $\alpha$ inside $F^{(3)}$), we have that the function $f$ is periodic in $s$, i.e., for every $m\in \mathbb{F}^b_2$ it holds that $f(m\oplus s)=f(m).$ The value $s$ can still be extracted by Simon's algorithm in quantum polynomial time $\mathcal{O}(b).$

\item Hence, the function $f$ can be used as an efficient quantum distinguisher. In general, the key extraction is possible in the case when the rolling function $roll_c$ is linear, in which case one applies \Cref{prop:systemK}.
\end{enumerate}
\begin{remark}
Note that similar construction and arguments can be applied to Farfalle-WBC-AE (given by \cite[Algorithm 6]{Farfale}), where one can manipulate the plaintext and/or metadata blocks (in terms of variables and constants).
\end{remark}

\section{Extracting a secret value from Simon's period in GFNs}\label{sec:period}

In many recent papers (for instance, see \cite{Cid,Palash,SHGFN3,Zhon,ZhouBV}), the application of Simon's or Simon-Grover algorithms to different key-alternating (Generalized) Feistel schemes provides a period of the form: 
$$s=F_k(\alpha)\oplus F_k(\beta),$$
where $\alpha$ and $\beta$ are known different constants, and $k$ is a secret round key. For instance, such an $s$ has been obtained in \cite{Hidenori} from the 3-round Feistel scheme (as shown in \Cref{ex:e1} earlier), Type 1, 2 and 3 GFNs in \cite{Dong,Dong2,SHSMS,SHGFN3,Ito1,ZhouBV}, etc. So far, no method has been proposed that extracts $k$. In this section we provide two methods for extracting the value $k$ (along with their formal descriptions, that is \Cref{alg:Keyextraction2} and \Cref{alg:Keyextraction}), when the given inner round function $F_k$ is defined as $F_k(z)=F(z\oplus k)$ by $F$ being publicly known function. Clearly, in the case when $F_k$ is defined as $F_k(z)=F(z)\oplus k$, then $s$ does not depend on $k$ at all. We note that these two methods are stemming from the same approach (presented below), while they only use different formulas for approximating an underlying function/polynomial. And finally, in \Cref{sec:lagrange2} we provide a generalization of the method presented in \Cref{sec:anfapproach} (along with the complexity evaluation), by employing high-order derivatives. Although the presented method imposes certain trade-offs between different involved parameters, it can be still viewed as an improvement of the method given in \Cref{sec:anfapproach}.

\subsection{General settings of the round-key extraction approach}\label{sec:gensec}

Let us assume that $F$ is a function in $n$-bits, i.e. $F:\mathbb{F}^n_2\rightarrow \mathbb{F}^n_2$, defined as  $F_k(z)=F(z\oplus k)$, $z\in \mathbb{F}^n_2.$ In order to extract the value $k$ from $s=F_k(\alpha)\oplus F_k(\beta)$, we firstly notice that the value of $s$ can be viewed as a function in $\alpha$ and $\beta$, i.e. we have that $s=s(\alpha,\beta).$ Recall that after some applications of Simon's algorithm (until one obtains $n-1$ linearly independent vectors orthogonal to $s$), we obtain only a particular value of $s$ for given $\alpha$ and $\beta$. Clearly, for higher values of $n$, it is not feasible to obtain all values of $s(\alpha,\beta)$ due to the large input space. As one can choose $\alpha$ and $\beta$ to be arbitrary and different, let us assume that: 
$$(\alpha,\beta)=(x,x\oplus \sigma),$$
where $\alpha=x$ and $\sigma\in \mathbb{F}^n_2$ is a non-zero fixed and known constant. Thus, we are considering the function $s:\mathbb{F}^n_2\rightarrow \mathbb{F}^n_2$ given by
$$s(x)=F_k(x)\oplus F_k(x\oplus \sigma),\;\;\;x\in \mathbb{F}^n_2.$$
Since $F$ is publicly known function, let us now consider the function $\Delta:\mathbb{F}^n_2\rightarrow \mathbb{F}^n_2$ defined by
$$\Delta(x)=s(x)\oplus F(x)\oplus F(x\oplus \sigma)=[F(x\oplus k)\oplus F(x)]\oplus [F(x\oplus \sigma\oplus k)\oplus F(x\oplus \sigma)].$$
Note that the function $\Delta$ has (at least) the non-trivial periods $\{k,\sigma,k\oplus \sigma\}$. The problem which remains to be solved is how to implement the function $\Delta$ efficiently in quantum/classical environment, without performing an infeasible amount of measurements (or applications of Simon's algorithm). In order to solve this problem, one may consider the following approaches.

\subsection{Utilizing an interpolation formula based on vector space representation}\label{sec:anfapproach}

In this subsection, we consider the inner function $F$ as a vectorial Boolean mapping from $\mathbb{F}^n_2 \rightarrow \mathbb{F}^n_2$. It is well-known that any Boolean function $g:\mathbb{F}^n_2 \rightarrow \mathbb{F}_2$ can be uniquely represented by its associated algebraic normal form (ANF) as follows:
\begin{eqnarray*}\label{ANF}
g(x_1,\ldots,x_n)={\bigoplus_{u\in \mathbb{F}^n_2}{\lambda_u}}{\left(\prod_{i=1}^n{x_i}^{u_i}\right)},
\end{eqnarray*}
where $x_i, \lambda_u \in \mathbb{F}_2$ and $u=(u_1, \ldots,u_n)\in \mathbb{F}^n_2$. The support of $g$ is defined as $supp(g)=\{x\in \mathbb{F}^n_2:g(x)=1\},$ and the algebraic degree of $g$ is defined as $deg_{alg}(g)=max\{wt(u):\lambda_u\neq 0\}$, where $wt(u)$ denotes the number of non-zero coordinates of $u\in \mathbb{F}^n_2.$

Let us now assume that the function $g:\mathbb{F}^n_2 \rightarrow \mathbb{F}_2$ has an algebraic degree $deg_{alg}(g)\leq d<n$, and that the values of $g$ are known on the set $S_d=\{y\in\mathbb{F}^n_2\;|\; wt(y)\leq d\}$. Then, according to \cite[see page 37]{CarletBoolean}, this can be used to recover correctly the whole function $g$ using the formula:
\begin{equation}\label{eq:recover}
g(x)=\bigoplus_{y\leq x, \; y\in S_d}g(y)\left[\left[\sum^{d-wt(y)}_{i=0}\binom{wt(x)-wt(y)}{i}\right][mod\;\; 2]\right].
\end{equation}
If we want to utilize (\ref{eq:recover}) in order to extract $k$ from $s(x)=F_k(x)\oplus F_k(x\oplus \sigma),$ $x\in \mathbb{F}^n_2,$ we consider the following setting. Recall that, for a vectorial Boolean function $F=(f_1,\ldots,f_n):\mathbb{F}^n_2\rightarrow \mathbb{F}^n_2$ with coordinate functions $f_i:\mathbb{F}^n_2\rightarrow \mathbb{F}_2$, we define the function $\lambda\cdot F:\mathbb{F}^n_2\rightarrow \mathbb{F}_2$ as $\lambda\cdot F(x)=\lambda_1f_1(x)\oplus \ldots \oplus \lambda_n f_n(x)$, with $\lambda=(\lambda_1,\ldots,\lambda_n)\in \mathbb{F}^n_2.$ Thus, for a given inner vectorial function $F$ and a non-zero vector $\lambda\in \mathbb{F}^n_{2}$, in what follows we will focus on the Boolean function $\lambda\cdot F$. Some additional discussions (conditions) related to the choice of $\lambda$ will be given later on.

In addition, by $d$ we denote the degree $d=deg_{alg}(\lambda\cdot F)$. We recall that the degree of any Boolean function $g:\mathbb{F}^n_2\rightarrow \mathbb{F}_2$ and its first derivative $D_{a}g(x)=g(x)\oplus g(x\oplus a)$ ($a\in \mathbb{F}^n_2$) are related by the inequality: 
$$deg_{alg}(g)\geq deg_{alg}(D_{a}g)+1.$$ 
Thus, we have that $t=deg_{alg}(\lambda\cdot s)=deg_{alg}D_{\sigma}(\lambda\cdot F)\leq d-1$, and consequently $\# S_t$ input-output pairs $(x_i,s(x_i))$ would be required in order to apply (\ref{eq:recover}), where $S_t=\{y\in\mathbb{F}^n_2\;|\; wt(y)\leq t=deg_{alg}(\lambda\cdot s)\}$. Clearly, knowing the pairs $(x_i,s(x_i))$ we can directly compute the values $\lambda\cdot s(x_i)$, and moreover, by $t\leq d-1$ we have:
$$\# S_t=\sum^{t}_{j=0}\binom{n}{j}\leq \sum^{d-1}_{j=0}\binom{n}{j}.$$
As $t=deg_{alg}(\lambda\cdot s)$ may not be known, then it is sufficient to require the amount of pairs equal to the sum on the right-hand side in the inequality given above. Note that one still uses the formula (\ref{eq:recover}), as it is allowed that the degree of $\lambda\cdot s$ is smaller than $d-1$.  Using (\ref{eq:recover}) we are able to obtain the function $\lambda\cdot s$ in classical environment, and thus we are able to implement the function $\Delta(x)=\lambda\cdot (s(x)\oplus F(x)\oplus F(x\oplus \sigma))$ in quantum environment and proceed with the extraction of $k$ by applying Simon's algorithm. 
\begin{remark}
Note that the choice of the vector $\lambda$ does not necessarily have to be the one which strictly minimizes the degree $deg_{alg}(\lambda\cdot F)$, i.e. it may be any non-zero vector and the presented method will still work (within reasonable complexity bounds). In general, the effect of the $deg_{alg}(\lambda\cdot F)$ is visible in \Cref{tab2} given later on. 
\end{remark}
We have the following result. 
\begin{proposition}\label{prop:period2}
For a vector $\lambda\in \mathbb{F}^n_{2}\setminus \{\textbf{0}_n\}$, let $d=deg_{alg}(\lambda\cdot F)$ be the algebraic degree of a publicly known function $F:\mathbb{F}^n_{2}\rightarrow \mathbb{F}^n_{2}$. Suppose that for all vectors $x\in S=\{y\in\mathbb{F}^n_2\;|\; wt(y)\leq d-1\}$ one knows the values of periods $s(x)=F(x\oplus k)\oplus F(x\oplus \sigma\oplus k)$ ($s:\mathbb{F}^n_{2}\rightarrow \mathbb{F}^n_{2}$, $\sigma\in \mathbb{F}^n_{2}\setminus \{\textbf{0}_n\}$) which are obtained by applying Simon's algorithm to some given function. Then:
\begin{enumerate}[1)]
\item One can recover the function $\lambda \cdot s(x)$ in classical environment without the knowledge of $k$ by using formula (\ref{eq:recover}).
\item By implementing the function $\Delta(x)=\lambda \cdot [s(x)\oplus F(x)\oplus F(x\oplus \sigma)]$ in quantum environment, one can recover $k$ by Simon's algorithm (provided that $\Delta$ is not a constant function).
\end{enumerate}
\end{proposition}
We point out the following properties related to \Cref{prop:period2}:
\begin{enumerate}[1)]
\item \Cref{prop:period2} loses its efficiency if the algebraic degree of $\lambda\cdot F(x)$ is high. For instance, the result can be used for some smaller values of $d=deg_{alg}(\lambda\cdot F)$, which is usually the case when $F$ represents an inner function of some GFN. \Cref{tab:compa} illustrates some scenarios regarding the parameters $d$ and $n$ (as the output size of $F$).

\item In the case when $\Delta(x)=\lambda\cdot (s(x)\oplus F(x)\oplus F(x\oplus \sigma))=\lambda\cdot D_{\sigma}D_{k}F(x)$ is a constant function, that is when $\lambda\cdot D_{\sigma}D_{k}F(x)=const.$ for all $x\in \mathbb{F}^n_2$, then clearly $\mathbb{F}^n_2$ is the space of periods of $\Delta$. 
The case when $\lambda\cdot D_{\sigma}D_{k}F(x)$ is a constant function, \textit{may occur} if $\lambda\cdot F(x)$ is at most $2$ or slightly higher (provided that the derivative directions $k$ and $\sigma$ are fast points\footnote{Let $g$ be a Boolean function. A fast point, say $\sigma$, is a vector for which $deg_{alg}(D_{\sigma}g)=deg_{alg}(g)-2.$}), or if $k=\sigma$ \cite[Theorem 2.15, pp. 45]{Wu}. Note that many of the existing GFNs are employing inner functions $F$ whose $deg_{alg}(\lambda\cdot F)$ can be taken to be easily higher than $3$ for many values of $\lambda$ (which occurs when $F$ employs larger S-boxes than 4 bits). In addition, $k=\sigma$ also has an extremely small probability, if $\sigma$ is taken uniformly at random (which can be imposed an assumption for the choice of $k$ as well).

\item In general, if the vectorial Boolean function $D_{\sigma}D_{k}F(x)$ is constant, then it necessarily holds that $F(x)\oplus F(x\oplus \sigma)=F(x\oplus k)\oplus F(x\oplus k\oplus \sigma)\oplus \widehat{C}$, for some constant $\widehat{C}\in \mathbb{F}^n_2.$ As we are considering that the inner function $F$  of a GFN has a strong structure, then its low differential uniformity (i.e. strong differential properties) does not admit this equality to hold for every $x\in \mathbb{F}^n_2.$ Therefore, if the interpolation of $\Delta(x)=\lambda\cdot D_{\sigma}D_{k}F(x)$  gives us a constant function, then one can simply consider instead the vectorial function $\Delta(x)=D_{\sigma}D_{k}F(x)$ (i.e. not its component $\lambda\cdot D_{\sigma}D_{k}F(x)$) and ensure that it is NOT a constant function due to its strong structure, unless some of the trivial cases happen, such as $k=\sigma$, or $F$ has low algebraic degree.

\item On the other hand, assuming that $\Delta$ is not a constant function in terms of the previous point, $k$ can be still extracted by performing sufficiently many measurements even when unwanted periods occur, due to \cite[Theorems 1 and 2]{Kaplan}.  In the case that $\Delta$ has no other periods than $\{k,\sigma,k\oplus \sigma\}$, which is highly expected, then the value $k$ can be deduced easily as the vector $\sigma$ is known ($k\neq\sigma$).
\end{enumerate}
\textbf{Addressing the case of $F=(f_1,\ldots,f_n)$ with $f_i$ having low algebraic degree:}  If for a given $\lambda\in \mathbb{F}^n_2$ we have that $\lambda\cdot F(x)$ has the algebraic degree equal to $3$, then $\Delta(x)=\lambda\cdot D_{\sigma}D_{k}F(x)$ will be a linear/affine Boolean function and thus having $2^{n-1}$ periods (which makes the search of $k$ infeasible for larger $n$). As some typical inner functions $F=(f_1,\ldots,f_n):\mathbb{F}^n_2\rightarrow \mathbb{F}^n_2$ in well-known GFNs today may have strong structure with coordinate functions $f_i$ having low algebraic degree, say $\leq 3$ (e.g. one may consider that $F$ is the round function of PRESENT block cipher \cite{Andrey}), then for the function $\Delta$ we can take a function with \textit{increased algebraic degree} instead of using the second derivative $D_{\sigma}D_{k}F(x)$. 

In order to provide the following result (whose proof is omitted due to its simplicity), we define the operation "$\ast$" as an operation between two arbitrary vectorial functions, say $G^{(i)}=(g^{(i)}_1,\ldots,g^{(i)}_n):\mathbb{F}^n_2\rightarrow \mathbb{F}^n_2$ ($i=1,2$), such that $G^{(1)}(x)\ast G^{(2)}(x)=(g^{(1)}_1(x)g^{(2)}_1(x),\ldots,g^{(1)}_n(x)g^{(2)}_n(x))$, i.e. we use the coordinate-wise products.
\begin{proposition}\label{prop:lowdeg}
Let $F=(f_1,\ldots,f_n):\mathbb{F}^n_2\rightarrow \mathbb{F}^n_2$ be a vectorial Boolean function. In addition, let $s(x,x\oplus \sigma)=F(x\oplus k)\oplus F(x\oplus k\oplus \sigma)$ be a period (viewed as a function) obtained from a given GFN (with respect to inputs $x,x\oplus \sigma\in \mathbb{F}^n_2$ in terms of \Cref{sec:gensec}). Then:
\begin{enumerate}[1)]
\item  The function $\Delta_1:\mathbb{F}^n_2\rightarrow \mathbb{F}^n_2$ given by
$$\Delta_1(x)=[F(x)\oplus F(x\oplus \sigma)]s(x,x\oplus \sigma)=[F(x)\oplus F(x\oplus \sigma)]\ast [F(x\oplus k)\oplus F(x\oplus k\oplus \sigma)]$$
has the periods $\{\textbf{0}_n,k,\sigma,k\oplus \sigma\}$.
\item  Define the function $S:\mathbb{F}^n_2\rightarrow \mathbb{F}^n_2$ by
$$S(x)=s(x,x\oplus \sigma_1)\oplus s(x,x\oplus \sigma_2)=F(x\oplus k\oplus \sigma_1)\oplus F(x\oplus k\oplus  \sigma_2),$$ where $\sigma_1\neq \sigma_2$ are two known non-zero vectors in $\mathbb{F}^n_2$, and $s(x,x\oplus \sigma_i)$ ($i=1,2$) are periods obtained from a given GFN. Then, the functions $\Delta_i:\mathbb{F}^n_2\rightarrow \mathbb{F}^n_2$ ($i=2,3$) given as
$$\begin{array}{c}
\Delta_2(x)=[F(x\oplus \sigma_1)\oplus F(x\oplus \sigma_2)]\ast S(x), \\ 
\Delta_3(x)=[F(x\oplus \sigma_1)\oplus F(x\oplus \sigma_2)]\oplus S(x),
\end{array} $$
have the periods $\{\textbf{0}_n,k,\sigma_1\oplus \sigma_2,k\oplus \sigma_1\oplus \sigma_2\}$.
\end{enumerate}
\end{proposition}
The main property of functions $\Delta_i$ in Proposition \ref{prop:lowdeg} is that they have either increased or decreased algebraic degree of its coordinate functions in comparison to $F$, as well as the function $D_{\sigma}D_{k}F(x)$ used in Proposition \ref{prop:period2}-$(2)$. In the case of Proposition \ref{prop:lowdeg}-$(1)$, we have that $\Delta_1(x)=[F(x)\oplus F(x\oplus \sigma)]s(x,x\oplus \sigma)=D_\sigma F(x)D_\sigma F(x\oplus k)$ has a "slightly" increased algebraic degree of coordinate functions in comparison to $F$. However, $\Delta_2(x)=[F(x\oplus \sigma_1)\oplus F(x\oplus \sigma_2)]\ast S(x)$ may have even higher algebraic degrees of coordinate functions in comparison to $F$, due to the fact that $F(x\oplus \sigma_1)\oplus F(x\oplus \sigma_2)$ is not a first-order derivative of $F$. Consequently, the constructions of $\Delta_i$ in Proposition \ref{prop:lowdeg} will not potentially suffer from being constant functions, as in the case of Proposition \ref{prop:period2}-$(2)$ which uses the construction $D_{\sigma}D_{k}F(x)$. Clearly, if the degree of the construction of $\Delta_2$ in Proposition \ref{prop:lowdeg}-$(2)$ is too high (due to the product operation "$\ast$"), then one may consider the function $\Delta_3$ which has a slightly reduced degree. Ultimately, as $\Delta_3(x)=D_k S(x)$ is the first derivative of $S(x),$ an algorithm that exploits high-order derivatives (by conveniently adjusting the degree) in order to extract $k$, is presented later in \Cref{sec:lagrange2}. 
\begin{remark}
Hence, Proposition \ref{prop:lowdeg} provides an efficient solution regarding the problem of potentially obtaining a constant function $\Delta$ in Proposition \ref{prop:period2}-$(2)$, as it provides constructions of functions (viewed in the place of $\Delta$) either with slightly increased or decreased algebraic degrees with respect to the coordinate functions of $F$. 
\end{remark}
Taking in consideration all the results given so far,  we conclude this subsection by providing \Cref{alg:Keyextraction2} which formally describes the round-key extraction method based on the relation (\ref{eq:recover}).\textit{ As a default version (for simplicity), we will use the construction of $\Delta$ given in Proposition \ref{prop:period2}-$(2)$}, while in the problematic case when  $F=(f_1,\ldots,f_n)$ has coordinate functions $f_i$ with low algebraic degrees, one may clearly utilize Proposition \ref{prop:lowdeg}. In that case, the most important fact here is that the complexity of the algorithm will not change significantly, as long as we do not consider the construction of $\Delta$ with high algebraic degree, due to the requirements of the interpolation formula (\ref{eq:recover}).
\begin{table}[h!]
\centering
\begin{tabular}{|c|c|c|c|c|c|}
\hline
$n\setminus d$ & $d=3$ & $d=4$ & $d=5$ & $d=6$ & $d=7$ \\ \hline
$n=32$ &$2^{9.04}$ & $2^{12.42}$ & $2^{15.34}$ & $2^{17.89}$ & $2^{20.13}$ \\ \hline
$n=64$ &$2^{11.02}$ & $2^{15.41}$ & $2^{19.37}$ & $2^{22.99}$ & $2^{26.31}$ \\  \hline
$n=128$ & $2^{13.01}$ & $2^{18.42}$ & $2^{23.39}$ & $2^{28.04}$ & $2^{32.40}$ \\  
\hline
\end{tabular} 
\vskip 2mm
\caption{Numbers of required input-output pairs $(x_i,\lambda\cdot s(x_i))$ needed to recover $\lambda \cdot s(x)$ by the formula (\ref{eq:recover}).}
\label{tab:compa}
\end{table}
%
%
%
%
%
\begin{algorithm}
  \renewcommand{\algorithmicrequire}{\textbf{Parameters:}}
  \caption{Round-key extraction based on the formula (\ref{eq:recover})} \label{alg:Keyextraction2}
  \begin{algorithmic}
    \Require Round function $F_k:\mathbb{F}^n_2\rightarrow \mathbb{F}^n_2$ of a block cipher is given by $F_k(z)=F(z\oplus k)$, $z\in \mathbb{F}^n_2$ ($k$ is a secret round key), with $d=deg_{alg}(\lambda\cdot F)\geq 4$, for suitable non-zero vector $\lambda\in \mathbb{F}^n_2$ (without expecting fast points of $\lambda\cdot F$, otherwise variate $\lambda$ or use Proposition \ref{prop:lowdeg}, cf. Remark \ref{rem:prop4}). The period $s$ (viewed as a function) is given as $s(x)=F_k(x)\oplus F_k(x\oplus \sigma)$ ($x\in \mathbb{F}^n_2$), where $\sigma \in \mathbb{F}^n_2$ is a non-zero fixed known constant.\\
    \State \textbf{Step 1 (On-line)} By applying classical Simon's algorithm, deduce the values of $s(x)$ for all vectors $x\in \{y\in\mathbb{F}^n_2\;|\; wt(y)\leq d-1\}$.
    \State \textbf{Step 2 (Off-line)} Recover the function $s(x)$ by applying the relation (\ref{eq:recover}). This is done in classical environment.
 \State \textbf{Step 2.1 (Off-line)} If the function $\Delta(x)=\lambda\cdot [s(x)\oplus F(x)\oplus F(x\oplus \sigma)]=\lambda\cdot D_{\sigma}D_k F(x)$ is constant, then  $k=\sigma$. Otherwise, for more confidence, consider the vectorial function $\Delta(x)=D_{\sigma}D_k F(x)$. Then go to the next step.
 
 \State \textbf{Step 3 (Off-line)} If $\Delta$ is not a constant function (whether we considered $\lambda\cdot D_{\sigma}D_k F(x)$ or $D_{\sigma}D_k F(x)$),  implement it quantum environment.
     
 \State \textbf{Step 4 (Off-line)}  Apply Simon's algorithm to $\Delta$ in order to obtain the set of periods $A=\{k,\sigma,k\oplus \sigma\}.$ 
     \State \textbf{Step 5 (Off-line)}  Deduce the value $k$ from $A$ (as $\sigma\in \mathbb{F}^n_2$ is known).
  \end{algorithmic}
\end{algorithm}
\begin{table}[h!]
\centering
\begin{tabular}{|c|c|c|c|}
\hline 
Attack model & Quantum queries & \makecell{Processing time\\ (In classical environment)} & Qubits used \\ 
\hline 
\textbf{Q2} & $\mathcal{O}(n\sum^{d-1}_{j=0}\binom{n}{j})$  & $n T+\mathcal{O}(n^3\sum^{d-1}_{j=0}\binom{n}{j})$ & $2n+2n\sum^{d-1}_{j=0}\binom{n}{j}$ \\ 
\hline 
\end{tabular} 
\vskip 2mm
\caption{Complexity estimates of \Cref{alg:Keyextraction2}. The parameter $T$ denotes the complexity of computing the expression in (\ref{eq:recover}). 
}
\label{tab2}
\end{table}
\begin{remark}\label{rem:prop4}
Note that the choice of $\Delta$ to be either $\lambda\cdot D_{\sigma}D_k F(x)$ or $D_{\sigma}D_k F(x)$ in \textbf{Step 3} (Algorithm \ref{alg:Keyextraction2}), does not require the repetition of \textbf{Steps 1} and \textbf{2}, since the full function $s(x)$ has been recovered. In this context, the corresponding interpolation processing complexity in Table \ref{tab2} is described by $nT$, since we are applying (\ref{eq:recover}) to all coordinate functions of $s(x).$ As discussed earlier, in order to avoid the case when $\Delta$ is a constant function in \textbf{Step 2.1} (happens when $F$ has coordinate functions of low degree), then one may simply consider the construction of $\Delta$ based on Proposition \ref{prop:lowdeg}, in which case \textbf{Steps 1} and \textbf{2} in Algorithm \ref{alg:Keyextraction2} stay the same with the parameter $d=max\{deg_{alg}f_i:F=(f_1,\ldots,f_n),\;i=1,\ldots,n\}$.
\end{remark}
\begin{remark}
Note that the processing time complexity in \Cref{tab2} encompasses  the Gaussian elimination procedure utilized for solving linear systems in Simon's algorithm (which corresponds to $\mathcal{O}(n^3)$). Also note that the last column represents the number of qubits used for the implementation of Simon's algorithm, where we count for both \textbf{Steps 1} and \textbf{4} ($2n$ qubits are used for a single implementation).
\end{remark}
\begin{remark}
We recall that the so-called off-line Simon's algorithm \cite{OfflineS} firstly requires (on-line) classical queries to a given encryption scheme (viewed as an oracle or black-box), and then applies (off-line) Grover's algorithm \cite{Grover} to find a  certain unknown value (where its classifying function is given in terms of Simon's algorithm). In the context of our approach, it is highly inefficient to require classical on-line queries in order to deduce sufficiently many periods $s(x_i)$ for arbitrary pairwise different inputs $x_i\in \mathbb{F}^n_2$. This is due to the fact that the extraction of a single period of a given function has exponential complexity (in terms of the total number of its inputs, and not a portion as in \cite{OfflineS}), and thus it does not admit efficient setting of the approach presented in \cite{OfflineS}. In addition, we note that our approach is utilizing Simon's algorithm only (multiple times), unlike the off-line Simon's algorithm \cite{OfflineS} which uses Grover's algorithm. We note that a similar reasoning applies to \Cref{alg:Keyextraction} given in the next subsection.
\end{remark}

\subsection{Utilizing the Lagrange interpolation formula}\label{sec:lagrange}

In this subsection, we consider the inner function $F$ as a univariate polynomial from $\mathbb{F}_{2^n} \rightarrow \mathbb{F}_{2^n}$, where $\mathbb{F}_{2^n}=GF(2^n)$ is a finite field of cardinality $2^n$. In order to recover the function $s:\mathbb{F}_{2^n} \rightarrow \mathbb{F}_{2^n}$, one may employ the so-called Lagrange interpolation formula (see e.g. \cite[Subsection 2.1.7.3]{Handbook}) which is described as follows. 

Let $x_1,\ldots,x_r,y_1,\ldots,y_r$ be elements of a field $\mathcal{F}$, where $x_i$, for $i\in\{1,2,\ldots,r\}$, are pairwise distinct. Then there exists a unique polynomial, say $h:\mathcal{F}\rightarrow \mathcal{F}$, with polynomial degree at most $r-1$ such that $h(x_i)=y_i$, $i\in \{1,2,\ldots,r\}$, which is given by
$$h(x)=\sum^r_{i=1}y_i\prod_{1\leq j\leq r,\;j\neq i}\frac{x-x_j}{x_i-x_j}.$$
\begin{remark}
Note that in the case when $F:\mathbb{F}_{2^n}\rightarrow \mathbb{F}_{2^n}$ ($\mathcal{F}=GF(2^n)=\mathbb{F}_{2^n}$) is given by its univariate representation in $\mathbb{F}_{2^n}[x]$ as $F(x)=\sum^{2^n-1}_{i=0}a_ix^i$, then its polynomial degree $deg_{poly}(F)$ is equal to maximal $i$ for which $a_i\neq 0$.
\end{remark}
%
%
%
If the publicly known function $F$ is of polynomial degree $d$, then $deg_{poly}(s)\leq deg_{poly}D_{\sigma}F(x) =d-1$ and thus one can interpolate the function $s$ if we have $(d-1)+1=d$ input-output pairs $(x_i,s(x_i)),$ $i=1,\ldots,d$. Note that this can be done even without knowing the secret value $k$. Consequently, by  Lagrange interpolation formula one can recover the function $\Delta:\mathbb{F}_{2^n}\rightarrow \mathbb{F}_{2^n}$, if we have $d$ input-output pairs $(x_i,s(x_i)),$ $i=1,\ldots,d$. We have the following result.
\begin{proposition}\label{prop:period}
Suppose that $s(x)=F(x+ k)+ F(x+ \sigma+ k)$ (viewed as $s:\mathbb{F}_{2^n}\rightarrow \mathbb{F}_{2^n}$, $\sigma\neq 0$ is known) is a period obtained by applying Simon's algorithm at input $x\in \mathbb{F}_{2^n}$ (in terms of \Cref{sec:gensec}), where $F:\mathbb{F}_{2^n}\rightarrow \mathbb{F}_{2^n}$ is a publicly known function with $deg_{poly}(F)=d$. In addition, assume that values $s(x_1),\dots,s(x_{d})$ have been extracted by taking any pairwise different inputs $x_1,\ldots,x_{d}\in \mathbb{F}_{2^n}$. Then:
\begin{enumerate}[1)]
\item One can recover $s(x)$ in classical environment without the knowledge of $k$, i.e., $$s(x)=\sum^{d}_{i=1}s(x_i)\prod_{1\leq j\leq n,\;j\neq i}\frac{x-x_j}{x_i-x_j}.$$
\item By implementing the function $\Delta(x)=s(x)\oplus F(x)\oplus F(x\oplus \sigma)$ (viewed as a vectorial Boolean mapping from $\mathbb{F}^n_{2}$ to $\mathbb{F}^n_{2}$) in quantum environment, one can recover $k$ by Simon's algorithm (provided that $\Delta$ is not a constant function).
\end{enumerate}
\end{proposition}
Note that in the case when $F$ is of higher degree, then the interpolation by Lagrange formula may become infeasible due to the large amount of required input-output pairs $(x_i,s(x_i))$. In this context, we recall the following lemma.
\begin{lemma}\cite{CCZ}\label{lema:deg}
Let $F:\mathbb{F}_{2^n}\rightarrow \mathbb{F}_{2^n}$ be a function and let $F(x)=\sum^{2^n-1}_{i=0}a_ix^i$ denote its univariate polynomial representation. The algebraic degree $deg_{alg}(F)=max\{deg_{alg}(\mathfrak{f}_i):F(x)=(\mathfrak{f}_1(x),\ldots,\mathfrak{f}_n(x)),\mathfrak{f}_i:\mathbb{F}^n_{2}\rightarrow \mathbb{F}_{2}\}$ of $F:\mathbb{F}^n_{2}\rightarrow \mathbb{F}^n_{2}$ viewed as a vectorial Boolean function is the maximum Hamming
weight of its exponents, i.e., $deg_{alg}(F) = max\{wt(i)| a_i\neq 0,\;i=0,1,\ldots,2^n-1\}$.
\end{lemma}
Hence, the main problem regarding the application of the Lagrange interpolation is that the polynomial degree $deg_{poly}(F)$ may be relatively high, although the algebraic degree $deg_{alg}(F)$ (defined as in \Cref{lema:deg}) is quite low. However, considering the fact that inner functions $F$ usually have low algebraic degree $deg_{alg}(F)$ (which is the case for many GFNs), \Cref{lema:deg} gives us the upper bound on the number of terms present in the polynomial representation of $F$ as follows. Namely, if $deg_{alg}(F)=\delta$, then the polynomial $F(x)=\sum^{2^n-1}_{i=0}a_ix^i$ has maximally
\begin{eqnarray}\label{eq:terms}
\sum^{\delta}_{j=0}\binom{n}{j}
\end{eqnarray}
non-zero terms in its representation, which is the total number of all possible terms $x^i$ with $i$ having weight $wt(i)\leq \delta.$ Considering that $\delta$ is not large enough (cf. also \Cref{tab:compa} for some values of $deg_{alg}(F)$), it means that $F$ can be considered as a sparse polynomial.\\\\
%
%
%
%
%
%
%
%
%
%
We conclude this subsection with the following remarks:
\begin{enumerate}[1)]
\item In \cite{Gong}, it has been pointed out that the choice of different irreducible polynomials used to construct the field $\mathbb{F}_{2^n}$ may affect the polynomial degree of a function defined on $\mathbb{F}_{2^n}$. In  this context, in order to reduce the number of required input-output pairs $(x_i,s(x_i))$ required in \Cref{prop:period}, one may firstly choose a suitable irreducible polynomial by which the polynomial degree of $F$ is minimized (which consequently minimizes $deg_{poly}(s)$). It has been concluded in \cite{Gong} that a linear transformation on the output coordinates affects the coefficients of the exponents that belong to the same cyclotomic cosets of the exponent in the original univariate function representation. On the other hand, a linear transformation on the input coordinates (or changing the irreducible polynomial) affects only the coefficients of the exponents with Hamming weight less than or equal to the maximum Hamming weight of the exponents in original univariate function representation.


\item For interpolating the polynomial $F(x)$ (having polynomial degree $d$) the complexity is $O(d \log d)$ and requires $d$ queries to the function. When $F$ is sparse and exact monomials present in $F$ is known, this complexity is reduced to $O(t \log t)$ where $t\ll d$ is the number of non-zero monomials present in $F$ (cf. the relation (\ref{eq:terms}) in the case when $deg_{poly}(F)=\delta$).

\item Note that one can apply quantum algorithm for interpolation particularly when degree $d$ of $F$ is large. An optimal quantum algorithm \cite{Chil} for interpolation requires $O(d/2)$ queries.
\end{enumerate}
The round-key extraction method (based on \Cref{prop:period}) and its complexity estimates, are given  as follows.
\begin{algorithm}
  \renewcommand{\algorithmicrequire}{\textbf{Parameters:}}
  \caption{Round-key extraction based on \Cref{prop:period}} \label{alg:Keyextraction}
  \begin{algorithmic}
    \Require Round function $F_k:\mathbb{F}^n_2\rightarrow \mathbb{F}^n_2$ of a block cipher is given by $F_k(z)=F(z\oplus k)$, $z\in \mathbb{F}^n_2$ ($k$ is a secret round key), with $deg_{poly}(F)=d$ being large enough such that $D_{\alpha}D_{\beta}F(x)$ is not constant regardless of $\alpha,\beta\in \mathbb{F}^n_{2}$ ($\alpha\neq \beta$); otherwise use \Cref{prop:lowdeg}. The period $s$ (viewed as a function) is given as $s(x)=F_k(x)\oplus F_k(x\oplus \sigma)$ ($x\in \mathbb{F}^n_2$), where $\sigma \in \mathbb{F}^n_2$ is a non-zero fixed known constant.\\
    \State \textbf{Step 1 (On-line)} By applying classical Simon's algorithm, deduce $d$ periods $s(x_1),\dots,s(x_{d})$ for arbitrary pairwise different inputs $x_1,\ldots,x_{d}\in \mathbb{F}^n_2$. 
    \State \textbf{Step 2 (Off-line)} Recover the function $s(x)$ by Lagrange interpolation formula (\Cref{prop:period}-$(1)$). This is done in classical environment.
   \State \textbf{Step 2.1 (Off-line)} If the function $\Delta(x)=[s(x)\oplus F(x)\oplus F(x\oplus \sigma)]$ is constant, then  $k=\sigma.$ For more confidence, one may use \Cref{prop:lowdeg} to construct a non-constant function $\Delta$. Otherwise, go to the next step.
     \State \textbf{Step 3 (Off-line)} Implement the function $\Delta(x)=s(x)\oplus F(x)\oplus F(x\oplus \sigma)$ ($x\in \mathbb{F}^n_2$) in quantum environment.
   \State \textbf{Step 4 (Off-line)}  Apply Simon's algorithm to $\Delta$ in order to obtain the set of periods $A=\{k,\sigma,k\oplus \sigma\}$.
     \State \textbf{Step 5 (Off-line)}  Deduce the value $k$ from $A$ (as $\sigma\in \mathbb{F}^n_2$ is known).
  \end{algorithmic}
\end{algorithm}
\begin{table}[h!]
\centering
\begin{tabular}{|c|c|c|c|}
\hline 
Attack model & Quantum queries & \makecell{Processing time\\ (In classical environment)} & Qubits used \\ 
\hline 
\textbf{Q2} & $\mathcal{O}(dn)$  & $\mathcal{O}(dn^3+d\cdot log(d))$ & $2n(d+1)$ \\ 
\hline 
\end{tabular} 
\caption{Complexity estimates of \Cref{alg:Keyextraction}.}
\label{tab1}
\end{table}
\begin{remark}
Note that in \Cref{tab1} the part $\mathcal{O}(d\cdot log(d))$ describes the complexity of computing Lagrange interpolation formula \cite{lagrangeCom} at $d$ points. 
\end{remark}

\subsection{Improving the method based on (\ref{eq:recover}) by imposing trade-offs}\label{sec:lagrange2}

In \Cref{sec:anfapproach} and \Cref{sec:lagrange}, we employ two different interpolation formulas in order to obtain the function $s(x)$. The secret round key is then extracted as a period of a (non-constant) function $\Delta(x)$ (whose construction uses $s(x)$). Depending on the representation of the function $s(x)$, whether it is a vector space or a finite field representation, we had different outcomes regarding the data complexity, i.e. different requirements on the number of input-output pairs $(x_i,s(x_i))$. Especially in the case of Lagrange's interpolation method (which uses finite fields representation), the potential obstacle may happen if the univariate representation of $s(x)$ contains very high polynomial degree (due to $F$), in which case \Cref{alg:Keyextraction} does not seem to be applicable. On the other hand, we have seen that the vector space representation is less affected, as algebraic degree of any component functions $\lambda\cdot F$ ($\lambda\in \mathbb{F}^n_2\setminus \{\textbf{0}_n\}$)  usually has a low value. 

In this subsection, we provide an improvement of the method based on the relation (\ref{eq:recover}) (presented in \Cref{sec:anfapproach}) by imposing certain trade-offs, such that one increases on-line quantum query complexity in terms of the number of applications of Simon's algorithm and consequently  reduces the interpolation data complexity  by constructing a suitable function with lower degree (which requires higher costs on the off-line implementation side). Later on, we discuss the main reasons which regard the impossibility of providing the same improvement for the method based on Lagrange interpolation (presented in \Cref{sec:lagrange}).

Before we provide a general approach related to the improvement of the method based on (\ref{eq:recover}), let us assume that the inner function
 $F$ is a mapping from $\mathbb{F}^n_2$ to $\mathbb{F}^n_2$, i.e. $F:\mathbb{F}^n_2\rightarrow \mathbb{F}^n_2$, since the same observations hold if one would consider a function $\lambda\cdot F$ for any $\lambda\in \mathbb{F}^n_2\setminus \{\textbf{0}_n\}.$ In the reminder of this subsection we will use the following definition.
\begin{definition}\label{def:algvec}
For a vectorial function $F:\mathbb{F}^n_2\rightarrow \mathbb{F}^n_2$ with coordinate functions $F(x)=(f_1(x),\ldots,f_n(x))$ ($f_i:\mathbb{F}^n_2\rightarrow \mathbb{F}_2$), we define its algebraic degree $deg_{alg}(F)$ by $deg_{alg}(F)=\max\{deg_{alg}f_i:F=(f_1,\ldots,f_i),\;i=1,\ldots,n\}$. 
\end{definition}
\textbf{General description of the method:} Firstly, we notice that the function $s(x)=F(x\oplus k)\oplus F(x\oplus k\oplus \sigma)$ ($x,\sigma\in \mathbb{F}^n_2$, $\sigma$ is a non-zero known fixed vector) is related to the first derivative of the function $F(x\oplus k)$ in the direction of $\sigma,$ i.e. $s(x)=D_{\sigma}F(x\oplus k)$. Furthermore, we have that the function $\Delta: \mathbb{F}^n_2\rightarrow \mathbb{F}^n_2$ can be written as
$$\Delta(x)=s(x)\oplus F(x)\oplus F(x\oplus \sigma)=D_{\sigma}D_{k}F(x),$$
which consequently gives that the degree of $\Delta$ (regardless of the representation) is at least by two less than $F$.

Now, \textit{the main observation in this context is the following:} Instead of interpolating the function $s(x)$ directly based on arbitrary input-output values $(x_i,s(x_i))$ with $x_i\neq x_j$ ($i\neq j$), we will use an increased amount of suitable pairs $(x_i,s(x_i\oplus \upsilon_i ,x_i\oplus \tau_i))$ for suitable $\upsilon_i,\tau_i\in \mathbb{F}^n_2$ in order to interpolate a high-order derivative $H(x)=D_{\sigma_\mu}\ldots D_{\sigma_1}D_k F(x)$. In this context, we will consider $s$ as a function in two arguments as $s(\alpha,\beta)=F_k(\alpha)\oplus F_k(\beta)$.\\\\
By this strategy, we will have an increased requirement on the number of suitable input-output pairs (which one obtains by applying Simon's algorithm), and decreased amount of required values $(x_i,H(x_i))$ for an interpolation (due to decreased degree of $H(x)$ in comparison to either $s(x)$ or $\Delta(x)$). Before we provide a formal description of the algorithm and complexity estimates, we consider the following computation which regards the function $H$.\\\\
\textbf{The main observations/properties:} Recall that in the off-line implementation of the function $\Delta(x)$ in \Cref{alg:Keyextraction2} and \Cref{alg:Keyextraction}, we use the publicly known function $F$ and the known vector $\sigma$, where we previously interpolate the function $s(x)$. In this context, by considering the function $s:\mathbb{F}^n_2\rightarrow \mathbb{F}^n_2$ as $s=s(x,x\oplus \sigma_1)$ ($\sigma_1\in \mathbb{F}^n_2\setminus \{\textbf{0}_n\}$), the function $H(x)=D_{\sigma_\mu}\ldots D_{\sigma_1}D_k F(x)$ can be written as follows:
\begin{eqnarray*}
H(x)&=&D_{\sigma_\mu}\ldots D_{\sigma_1}D_k F(x)=(D_{\sigma_\mu}\ldots D_{\sigma_1})[F(x)\oplus F(x\oplus k)]\\
&=&D_{\sigma_\mu}\ldots D_{\sigma_2}D_{\sigma_1}F(x)\oplus D_{\sigma_\mu}\ldots D_{\sigma_2}[F(x\oplus k)\oplus F(x\oplus k\oplus \sigma_1)]\\
&=&D_{\sigma_\mu}\ldots D_{\sigma_2}D_{\sigma_1}F(x)\oplus D_{\sigma_\mu}\ldots D_{\sigma_2}s(x,x\oplus \sigma_1).
\end{eqnarray*}
From this expression, we conclude the following facts:
\begin{enumerate}[1)]

\item In order to interpolate $H(x)$ (by any of the methods given in \Cref{sec:anfapproach} and \Cref{sec:lagrange}), firstly we need to acquire the pairs $(x_i,s(x_i\oplus \upsilon,x_i\oplus \sigma_1\oplus \upsilon))$ for all $\upsilon\in \Phi=\langle \sigma_2,\ldots,\sigma_\mu\rangle=\{c_1\sigma_2\oplus \ldots\oplus c_{\mu-1}\sigma_\mu:(c_1,\ldots, c_{\mu-1})\in \mathbb{F}^{\mu-1}_2\}$ for pairwise different $x_i$, due to the term $D_{\sigma_\mu}\ldots D_{\sigma_2}s(x,x\oplus \sigma_1).$ \textit{For every fixed} $x_i\in \mathbb{F}^n_2$, these pairs are obtained by applying Simon's algorithm to the given block cipher $2^{\mu-1}$ times, which corresponds to the cardinality of $\Phi.$ Note that we do not specify the number of required input-output pairs  $(x_i,s(x_i\oplus \upsilon,x_i\oplus \sigma_1\oplus \upsilon))$, as it depends on the interpolation method. An example for the expression $D_{\sigma_3}D_{\sigma_2}s(x,x\oplus \sigma_1)$ is:
\begin{eqnarray*}
D_{\sigma_3}D_{\sigma_2}s(x,x\oplus \sigma_1)&=&s(x,x\oplus \sigma_1)\oplus [s(x\oplus \sigma_2,x\oplus \sigma_1\oplus \sigma_2)]\\
&&\oplus [s(x\oplus \sigma_3,x\oplus \sigma_1\oplus \sigma_3)]\\
&&\oplus [s(x\oplus \sigma_2 \oplus \sigma_3,x\oplus \sigma_1\oplus \sigma_2 \oplus \sigma_3)],
\end{eqnarray*}
and thus $D_{\sigma_3}D_{\sigma_2}s(x,x\oplus \sigma_1)$ contains the terms $(x,s(x\oplus \upsilon,x\oplus \sigma \upsilon))$, where $\upsilon$ belongs to the space spanned by $\sigma_2$ and $\sigma_3$, that is $\upsilon\in\{c_1\sigma_2\oplus c_2\sigma_3:(c_1,c_2)\in\mathbb{F}^2_2\}$.

\item Furthermore, for an interpolation we need to compute the values $(x_i,H(x_i))$. For this part, we firstly need to evaluate the function $D_{\sigma_\mu}\ldots D_{\sigma_2}D_{\sigma_1}F(x)$ in $x_i$ ($F$ and $\sigma_1,\ldots,\sigma_\mu$ are known). In addition, one has to construct the expression $D_{\sigma_\mu}\ldots D_{\sigma_2}s(x_i,x_i\oplus \sigma_1)$ by using the values $s(x_i\oplus \upsilon,x_i\oplus \sigma_1\oplus \upsilon)$, $\upsilon\in \Phi$.

\item Clearly, the space of vectors spanned by the set $\{\sigma_1,\ldots,\sigma_\mu,k\}$ is actually the space of periods of the function $H(x)=D_{\sigma_\mu}\ldots D_{\sigma_1}D_k F(x)$. However, it is well-known that if $\sigma_1,\ldots,\sigma_\mu,k$ are linearly dependent, then $H(x)=D_{\sigma_\mu}\ldots D_{\sigma_1}D_k F(x)$ is the zero-function \cite[Theorem 2.15, pp. 45]{Wu}. Consequently, if we consider linearly independent vectors $\sigma_1,\ldots,\sigma_\mu\in \mathbb{F}^n_2\setminus \{\textbf{0}_n\}$, then the case when $k\in \Sigma=\langle \sigma_1,\ldots,\sigma_\mu \rangle
=\{\bigoplus^{\mu}_{i=1}c_i\sigma_i:c_i\in \mathbb{F}_2\}$ implies that $H$ is the zero-function (i.e. $H(x)=\textbf{0}_n$ for all $x\in \mathbb{F}^n_2$). Here we obtain the bound $\mu\leq n-1$, since $\mu=n$ gives that $\Sigma=\mathbb{F}^n_2$ and thus $H$ is the zero-function.

\item If $deg_{alg}(F)=d$ (cf. \Cref{def:algvec}), then $deg_{poly}(H)\leq d-(\mu+1)$. Thus, this bound on $deg_{alg}(H)$ gives weaker requirements on the number of considered input-output values $(x_i,H(x_i))$ for an interpolation based on (\ref{eq:recover}). Here clearly, the case in which the choice of $\mu$ does not admit the extraction of $k$ is when $\mu\geq d-1,$ since in this case $deg_{alg}(H)=0$ (regardless of $F$) and thus no information can be gained in general. Hence, the general reasonable bounds on $\mu$ are given by $1\leq \mu \leq d-2$.

\item The main obstacle that we encounter regarding the finite field representation, is that we do not know how high-order derivatives affect directly the polynomial degree $deg_{poly}(F)$, i.e. what is the direct relation between the degree of $deg_{poly}(H)$ in terms of $deg_{poly}(F)$. A naive approach would be to derive explicitly a univariate representation of $H$ (and then read its polynomial degree), in which case we may know the data complexity required for the Lagrange interpolation. Otherwise, to the best of our knowledge, we could not find a result which can directly describe the relation between the polynomial degrees of $H$ and $F$. For this reason, we are only presenting an approach which regards the method based on the vector space representation given in \Cref{sec:anfapproach}.

\item Suppose that $H$ is not the zero-function, i.e. when $k\not\in \Sigma$ and $\sigma_1,\ldots,\sigma_\mu$ are linearly independent. Then clearly with very high probability $H$ has $\mu+1$ periods, and thus the whole space of periods $A=\Sigma\cup (k\oplus \Sigma)$ can be generated by any of its $\mu+1$ linearly independent elements. In other words, when extracting the periods of $H$, we only need $\mu+1$ linearly independent periods (since they span the whole space $A$). 

\item In general, whenever in the remaining of this subsection we discuss the cases whether $H$ is constant or not, we will consider it in terms of the linear dependency between the underlying direction vectors $\sigma_i$ and $k$, due to the strong structure of the function $F$ (no fast points are expected).

\item Regarding the use of interpolation method presented in \Cref{sec:anfapproach} (based on (\ref{eq:recover})), in the place of the function $H$ we may consider either  one of its components $\lambda\cdot H$ similarly as in \Cref{alg:Keyextraction2} (for some $\lambda\cdot \mathbb{F}^n_2\setminus \{\textbf{0}_n\}$), or we can apply (\ref{eq:recover}) to every of its Boolean coordinate functions $h_i$, where $H=(h_1,\ldots,h_n)$. If we want to apply (\ref{eq:recover}) in order to recover $H$ as a function from $\mathbb{F}^n_2\rightarrow \mathbb{F}^n_2$, then recovering of every $h_i$ will still utilize at maximum $\sum^{d-(\mu+1)}_{j=0}\binom{n}{j}$ input-output pairs $(x_i,H(x_i))$ due to $deg_{alg}(H)\leq d-(\mu+1)$, where $d=deg_{alg}(F)$ (cf. \Cref{def:algvec}). Then clearly, if $T$ represents the complexity of interpolating a single function $h_i$ by applying (\ref{eq:recover}), then $nT$ will correspond  to the interpolation of $n$ functions $h_1,\ldots,h_n.$

\item And finally, we note that the function $H$ can also be constructed as a high-order derivative of some $\Delta_i$ function given in \Cref{prop:lowdeg}. As the constructions of $\Delta_i$ may have different degrees than $F$, it consequently gives different bounds for the parameter $\mu$ (which affects the order of derivative used in $H$). Another useful property would be that $H$ may not be a constant function after the interpolation, due to the fact that by suitable high-order derivatives one may reduce the degree of a considered $\Delta_i$ function in a convenient way. Clearly, different constructions of $H$ are imposing different trade-offs and complexities, whose analyses are outside of the scope of this work.
\end{enumerate}
On the other hand, the extraction of $k$ can be done efficiently as follows. Firstly recall that the space of periods of $H(x)=D_{\sigma_\mu}\ldots D_{\sigma_1}D_k F(x)$  is spanned by the set of vectors $\{\sigma_1,\ldots,\sigma_\mu,k\}$. In order to extract $k$, whether $k$ is dependent or independent from $\sigma_1,\ldots,\sigma_\mu$ (which is not known in advance), one can take an advantage of choosing suitable vectors $\sigma_i$. More precisely, \textit{the main idea in this context is to construct suitable different high-order derivatives of $F$}, that is, functions 
$H^{(j)}(x)=D_{\sigma^{(j)}_\mu}\ldots D_{\sigma^{(j)}_1}D_k F(x)$ which will utilize suitable sets of vectors $\sigma^{(j)}_1,\ldots,\sigma^{(j)}_\mu\in \mathbb{F}^n_2$, for $j=1,2,\ldots,\lfloor\frac{n}{n-\mu}+1\rfloor$ in order to recover all bits of $k$ (cf. Remark \ref{rem:bound} given later on). 
We illustrate our approach and the main idea throughout the following example.
\begin{example}\label{ex:k}
Let us assume that the inner function $F$, defined as $F_k(z)=F(z\oplus k)$, is a mapping from $\mathbb{F}^{32}_2$ to $\mathbb{F}^{32}_2$ (thus $n=32$). We will consider the algebraic case of the interpolation (i.e. the use of (\ref{eq:recover})), and thus assume further that the algebraic degree of a component function $\lambda\cdot F$ (for some $\lambda\in \mathbb{F}^{32}_2\setminus \{\textbf{0}_{32}\}$) is equal to $d=deg_{alg}(\lambda\cdot F)=8$. Furthermore, suppose that we want to extract $k$ by constructing higher-order derivatives. In this context, we are free to choose the parameter $1\leq \mu\leq d-2=6$, which determines the order of derivatives that we want to construct. In order to recover the full round key $k$, let us consider the value $\mu=3$, and construct the following high order derivatives:\\\\
\textit{Step 1:} Let us consider the vectors $\sigma^{(1)}_1,\sigma^{(1)}_2,\sigma^{(1)}_3\in \mathbb{F}^{32}_2$ given by $\sigma^{(1)}_1=(1,0,0,$ $\textbf{0}_{n-\mu})=(1,0,0,\textbf{0}_{29})$,  $\sigma^{(1)}_2=(0,1,0,\textbf{0}_{29})$ and $\sigma^{(1)}_2=(0,0,1,\textbf{0}_{29})$. The main property of these three vectors is that their last $n-\mu=29$ coordinates are equal to zero, while in the remaining part (i.e., in the fist $\mu=3$) coordinates they have only one non-zero coordinate. Now, consider the derivative
$$H^{(1)}(x)=D_{\sigma^{(1)}_3}D_{\sigma^{(1)}_2}D_{\sigma^{(1)}_1}D_k  F(x),$$
and apply the procedure explained earlier to the function $\lambda\cdot H^{(1)}(x)=D_{\sigma^{(1)}_3}D_{\sigma^{(1)}_2}$ $D_{\sigma^{(1)}_1}D_k (\lambda \cdot F(x))$, i.e. query for suitable values $(x_i,s(x_i\oplus v,x_i\oplus \sigma^{(1)}_1\oplus  v))$, $v\in \Phi=\langle  \sigma^{(1)}_2, \sigma^{(1)}_3\rangle$ and interpolate $\lambda\cdot H^{(1)}$ by (\ref{eq:recover}) (note that $deg_{alg}(\lambda\cdot H^{(1)})\leq d-(\mu+1)=8-(3+1)=4$). Now, we come to the part which is related to the application of Simon's algorithm to $\lambda\cdot H^{(1)}$ in order to extract the periods spanned by the set $\{\sigma^{(1)}_1,\sigma^{(1)}_2, \sigma^{(1)}_3,k\}$. We distinguish the following cases:\\\\
\textbf{The case when $\lambda\cdot H^{(1)}\neq 0$ ($k\not \in \Sigma$):}  The space of periods of $\lambda\cdot H^{(1)}$ in this case is $A=\Sigma\cup (k\oplus \Sigma)$. Now, it is clear that the only part of $A$ which is containing the key bits is its affine subspace $k\oplus \Sigma\subset A$. Thus, let us consider an arbitrary vector $\gamma\in k\oplus \Sigma\subset A$, which has the form $\gamma=k\oplus \sigma_{\gamma}=(\gamma_1,\ldots,\gamma_{32})$, for some $\sigma_{\gamma}\in \Sigma.$ Note that whether $\sigma_{\gamma}$ is equal to $\textbf{0}_{32}$ or not, it is not known. Hence, the main goal here is to recover $k=(\kappa_1,\ldots,\kappa_{32})\in \mathbb{F}^{32}_2$, where $\gamma$ is known (as a period of $\lambda\cdot H^{(1)}$) and $\sigma_{\gamma}$ is not known. Using the fact that the last $n-\mu=29$ coordinates of $\sigma^{(1)}_i$ are equal to zero, $i=1,2,3$ (as is the case for all vectors in $\Sigma=\langle \sigma^{(1)}_1, \sigma^{(1)}_2, \sigma^{(1)}_3\rangle$), we clearly have that the last $29$ coordinates of $\gamma$ actually represent the last $29$ coordinates of $k$, i.e.  $k=(\kappa_1,\kappa_{2},\kappa_{3},\gamma_4,\ldots,\gamma_{32})\in \mathbb{F}^n_2$.\\\\
\textbf{The case when $\lambda\cdot H^{(1)}= 0$ ($k\in \Sigma$):} In this case, $k$ is a linear combination of $\sigma^{(1)}_1, \sigma^{(1)}_2, \sigma^{(1)}_3$. As  the last $n-\mu=29$ coordinates of $\sigma^{(1)}_i$ are equal to zero, then clearly it holds that $k$ is given as $k=(\kappa_1,\kappa_{2},\kappa_{3},\textbf{0}_{29})$.\\\\
\textit{Step 2:} Previously, we have seen that after the interpolation of $\lambda\cdot H^{(1)}$, whether we get that $\lambda\cdot H^{(1)}$ is the zero-function or not, we still can recover $n-\mu$ coordinates of $k$. In order to recover the fist $\mu=3$ coordinates of $k$, we simply choose new vectors $\sigma^{(2)}_i$ ($i=1,2,3$) such that their first $n-\mu=29$ coordinates are equal to $\textbf{0}_{29}$, and the remaining parts contain just one non-zero coordinate. In other words, we consider the vectors $\sigma^{(2)}_1=(\textbf{0}_{29},1,0,0)$,  $\sigma^{(2)}_1=(\textbf{0}_{29},0,1,0)$ and $\sigma^{(2)}_1=(\textbf{0}_{29},0,0,1)$. After applying the same procedure as in \textit{Step 1} to the function
$$\lambda\cdot H^{(2)}(x)=D_{\sigma^{(2)}_3}D_{\sigma^{(2)}_2}D_{\sigma^{(2)}_1}D_k(\lambda\cdot F(x)),$$
we will be able to recover the first $n-\mu=29$ coordinates of $k$, regardless of whether $\lambda\cdot H^{(2)}$ is the zero function or not. At this point we remark the following possibilities with respect to \textit{Step 1}. Assume that in \textit{Step 1} we had that $k=(\kappa_1,\kappa_{2},\kappa_{3},\textbf{0}_{29})$, then in \textit{Step 2} it can not happen that $k$ is a linear combination of $\sigma^{(2)}_i$, in which case we would get a contradiction (as the last $29$ coordinates of $k$ are equal to $0$). On the other hand, the case $\lambda\cdot H^{(1)}\neq 0$ in \textit{Step 1} does not impose contradictions to any of the possibilities in \textit{Steps 2} (i.e. to any of $\lambda\cdot H^{(2)}\neq 0$ or $\lambda\cdot H^{(2)}= 0$). 

And finally, we point out that one could choose other high order derivatives in the place of $\lambda\cdot H^{(2)}$, in order to recover the first $3$ bits of $k$. For instance, one could choose $\mu=4$ vectors $\sigma^{(2)}_1,\ldots,\sigma^{(2)}_4$ whose first $28$ coordinates are equal to zero, and the remaining parts contain only one non-zero coordinate. Still, we do not violate the bound on $\mu$ given in (iv) by $1\leq \mu\leq d-2=8-2=6$.

Overall, we needed at maximum $\lfloor\frac{n}{n-\mu}+1\rfloor=\lfloor\frac{32}{32-3}+1\rfloor=2$ high-order derivatives in order to recover all bits of $k$. At the same time, this represents the upper bound on the number of times that we have to apply the interpolation procedure (along with the requirement on input-output pairs) and extract periods by Simon's algorithm. \qed
\end{example}
\begin{remark}\label{rem:bound}
Recall that per each step above, one recovers $n-\mu$ bits of $k$, regardless of whether $k\in \Sigma$ or $k\not\in \Sigma$. Thus, one has to proceed with at least $\lfloor\frac{n}{n-\mu}\rfloor$ steps in order to recover most of the bits of $k$, if not all. In the case when $n-\mu$ divides $n$, then the exact number of steps in order to recover all bits of $k$ is precisely $\frac{n}{n-\mu}=\lfloor\frac{n}{n-\mu}\rfloor$. Otherwise, when $n-\mu$ does not divides $n$, one clearly needs $\lfloor\frac{n}{n-\mu}+1\rfloor$ steps to recover all bits of $k$ as explained in Example \ref{ex:k}. In any case, $\lfloor\frac{n}{n-\mu}+1\rfloor$ is the maximal number of steps needed in order to recover all bits of $k$, which we consider in the detailed explanation below (as well as in \textbf{Parts I} and \textbf{II} in Algorithm \ref{alg:Keyextraction3} later on).
\end{remark}
%
%
%
In what follows, we provide a formal description of the observations related to $H$ given earlier and the method presented in \Cref{ex:k}, that is \Cref{alg:Keyextraction3} along with its complexity estimates in \Cref{tab3}. We note that \Cref{alg:Keyextraction3} already incorporates the key extraction method presented in \Cref{ex:k}, which overall gives the factor $J=\lfloor\frac{n}{n-\mu}+1\rfloor$ throughout all complexities in \Cref{tab3}. In addition, \Cref{tab4}  illustrates some cases of relevant parameters in terms of \Cref{alg:Keyextraction3}.\\\\
\textbf{Notation for \Cref{alg:Keyextraction3}:} We will use the notation of canonical vectors, that is, by  $e^{(i)}$ we denote the vector $e^{(i)}=(0,\ldots,1,\ldots,0)\in \mathbb{F}^{\mu}_2$, where the value $1$ stands at $i$-th coordinate, $i=1,\ldots,\mu$. Note that $\mu=1$ means that we are considering $H(x)=D_{\sigma_1}D_kF(x)$, which is actually the function $\Delta$ in \Cref{sec:anfapproach} and \Cref{sec:lagrange}. \\\\
\begin{algorithm}
  \renewcommand{\algorithmicrequire}{\textbf{Parameters:}}
  \caption{Round-key extraction based on High-order derivatives} \label{alg:Keyextraction3}
  \begin{algorithmic}
    \Require An inner round function $F_k:\mathbb{F}^n_2\rightarrow \mathbb{F}^n_2$ of a given GFN block cipher defined by $F_k(z)=F(z\oplus k)$, $z\in \mathbb{F}^n_2$ ($k$ is a secret round key), with $deg_{alg}(F)=d$. The period $s$ (viewed as a function) is given in general as $s(x,x\oplus \sigma)=F_k(x)\oplus F_k(x\oplus \sigma)$ ($x\in \mathbb{F}^n_2$), where in the place of fixed constant $\sigma \in \mathbb{F}^n_2\setminus \{\textbf{0}_n\}$ we will consider the vectors $\sigma^{(j)}_1$ (in \textbf{Part I}) and $\sigma_1$ (in \textbf{Part II}) defined below.\\\\
\textbf{Part I:} Fix a value $\mu\in\{1,\ldots,d-2\}$. For $j=1,2,\ldots,\lfloor\frac{n}{n-\mu}\rfloor$, proceed with the following steps:

\State \textbf{Step 0 (Off-line)} Prepare pairwise distinct inputs $x_1,\ldots,x_{\varepsilon}\in \mathbb{F}^n_2$, and define vectors $\sigma^{(j)}_i=(s_1,\ldots,s_n)\in  \mathbb{F}^n_2\setminus \{0\}_n$, $i=1,\dots,\mu,$ by 
$$\left\{\begin{array}{cc}
(s_{n-j(n-\mu)+1},\ldots,s_{n-(j-1)(n-\mu)})=\textbf{0}_{n-\mu}, &  \\ 
(1,\ldots,s_{n-j(n-\mu)},s_{n-(j-1)(n-\mu)+1},\ldots,s_n)=e^{(i)}, & 1\leq n-j(n-\mu)\leq \mu-1, \\ 
(s_{n-(j-1)(n-\mu)+1},\ldots,s_n)=e^{(i)}, & n-j(n-\mu)<1, \\ 
(1,\ldots,s_{n-j(n-\mu)})=e^{(i)}, & (j-1)(n-\mu)<1,
\end{array} \right.$$
where $e^{(i)}\in \mathbb{F}^{\mu}_2.$

\State \textbf{Step 1 (On-line)}  By applying Simon's algorithm, deduce periods $s(x_i\oplus \upsilon,x_i\oplus \sigma^{(j)}_1\oplus \upsilon)$ for all $\upsilon\in \Phi=\{c_1\sigma^{(j)}_2\oplus \ldots\oplus c_{\mu-1}\sigma^{(j)}_\mu :(c_1,\ldots, c_{\mu-1})\in \mathbb{F}^{\mu-1}_2\}$.


    \State \textbf{Step 2 (Off-line)}  By using the pairs $(x_i,s(x_i\oplus \upsilon,x_i\oplus \sigma^{(j)}_1\oplus \upsilon))$, $\upsilon\in \Phi$, compute $H^{(j)}(x_i)=D_{\sigma^{(j)}_\mu}\ldots D_{\sigma^{(j)}_1}F(x_i)\oplus D_{\sigma^{(j)}_\mu}\ldots D_{\sigma^{(j)}_2}s(x_i,x_i\oplus \sigma^{(j)}_1)$, $i=1,\ldots,\varepsilon.$  This part is done in classical environment. 

 \State    \textbf{Step 3 (Off-line)} Using the pairs $(x_i, H^{(j)}(x_i))$, $i=1,\ldots,\varepsilon,$  interpolate the function $G=\lambda\cdot H^{(j)}:\mathbb{F}^n_2\rightarrow \mathbb{F}_2$ using the relation (\ref{eq:recover}), for some fixed $\lambda\in \mathbb{F}^n_2\setminus \{\textbf{0}_n\}$.
 %
The interpolation is done in classical environment. 
 \State    \textbf{Step 3.1 (Off-line)}  If $G$ is the zero-function, then $k=(\kappa_1,\ldots,\kappa_n)\in \Sigma=\{\bigoplus^{\mu}_{i=1}c_i\sigma^{(j)}_i:c_i\in \mathbb{F}_2\}$ and thus $(\kappa_{n-j(n-\mu)+1},\ldots,\kappa_{n-(j-1)(n-\mu)})=\textbf{0}_{n-\mu}.$
     \State \textbf{Step 4 (Off-line)} If $G$ is NOT the zero-function, implement the function $G$ in quantum environment. Then, apply Simon's algorithm in order to obtain the space of its periods, that is $A=\Sigma \cup (k\oplus \Sigma)$.
     \State \textbf{Step 6 (Off-line)} For an arbitrary $\gamma=(\gamma_1,\ldots,\gamma_n)\in A$ with $(\gamma_{n-j(n-\mu)+1},\ldots,\gamma_{n-(j-1)(n-\mu)})\neq \textbf{0}_{n-\mu}$, we obtain $(\kappa_{n-j(n-\mu)+1},\ldots,\kappa_{n-(j-1)(n-\mu)})=(\gamma_{n-j(n-\mu)+1},\ldots,\gamma_{n-(j-1)(n-\mu)})$. 
   
   \State \textbf{Output of Part I:}  If $(n-\mu)|n$ and one proceeds with all the steps above for all  $j=1,2,\ldots,\lfloor\frac{n}{n-\mu}\rfloor$, then one obtains the whole key $k$.   
     \\\\
     \textbf{Part II:} If $(n-\mu)\not |n$, then after applying \textbf{Part I} define new vectors $\sigma_i\in \mathbb{F}^n_2\setminus \{0_n\}$ ($i=1,\ldots,\mu$)  by $\sigma_i=(\textbf{0}_{n-\mu},e^{(i)})$, $i=1,\dots,\mu.$ By obtaining the periods $(x_i,s(x_i\oplus \upsilon,x_i\oplus \sigma_1\oplus \upsilon))$, $\upsilon\in \langle \sigma_1,\ldots,\sigma_\mu\rangle$, obtain the values $(x_i,H(x_i))$, where $H(x)=D_{\sigma_\mu}\ldots D_{\sigma_1}D_k F(x)$. After interpolating $G=\lambda\cdot H:\mathbb{F}^n_{2}\rightarrow \mathbb{F}_{2}$ by (\ref{eq:recover}) (for some non-zero $\lambda\in\mathbb{F}^n_2$), and applying Simon's algorithm,\textbf{ deduce the first $n-\mu$ bits of $k$} from the space of periods $A$ similarly as in \textbf{Part I} (cf. \Cref{ex:k}).
  \end{algorithmic}
\end{algorithm}
%
%
%
%

\begin{table}[h!]
\scriptsize
\centering
\begin{tabular}{|c|c|c|c|}
\hline 
Attack model \textbf{Q2} & Quantum queries & \makecell{Processing time \\(In classical environment)} & Qubits used \\ 
\hline 
\makecell{In general} & $\mathcal{O}(nJ\varepsilon 2^{\mu-1})$  & $J[Comp_{Int}+ \mathcal{O}(n^3J\varepsilon 2^{\mu-1})$] & $\displaystyle  J[2n+2n\varepsilon2^{\mu-1}]$ \\  \hline 
\makecell{Interpolation via \\ the relation (\ref{eq:recover})} & $\displaystyle  \mathcal{O}(nJ2^{\mu-1}\sum^{d-(\mu+1)}_{j=0}\binom{n}{j})$  & $JT+ \displaystyle \mathcal{O}(n^3J2^{\mu-1}\sum^{d-(\mu+1)}_{j=0}\binom{n}{j})$ & \makecell{$2nJ+$\\ $\displaystyle  2nJ 2^{\mu-1}\sum^{d-(\mu+1)}_{j=0}\binom{n}{j}$ }\\ 
\hline 
\end{tabular} 
\caption{Complexity estimates of \Cref{alg:Keyextraction3}.  The parameter $J=\lfloor\frac{n}{n-\mu}+1\rfloor$ and $T$ denote the complexity of computing the sum in (\ref{eq:recover}) with $\sum^{d-(\mu+1)}_{j=0}\binom{n}{j}$ addends. Also, $Comp_{Int}$ denotes the processing complexity of underlying interpolation.}
\label{tab3}
\end{table}
%
%
%
%
\begin{table}[h!]
\centering
\begin{tabular}{|c|c|c|c|c|c|}
\hline 
\makecell{Parameters \\ $n,d,J$} & $\mu$ & \makecell{Quantum\\ queries} & \makecell{Processing time \\(Classical\\ environment)} & \makecell{Interpolation \\ data complexity\\ $J\sum^{d-(\mu+1)}_{j=0}\binom{n}{j}$} &  \makecell{Qubits\\ used}  \\ 
\hline 
\multirow{2}{*}{\makecell{$n=16$, $d=4$\\($J=2$)}} & $\mu=1$ & $2^{12.1}$   & $2T+2^{20.1}$ & $2^{12.29}$  &  $8832$  \\  \cline{2-6} 
& $\mu=2$ &  $2^{10.1}$    & $2T+2^{18.1}$ & $2^{10.45}$  & $2240$ \\  \hline
\multirow{2}{*}{\makecell{$n=32$, $d=6$\\($J=2$)}} & $\mu=3$ &  $2^{17.1}$    & $2T+2^{27}$ &  $2^{16.34}$ & $270976$\\  \cline{2-6} 
& $\mu=4$ &    $2^{14}$    & $2T+2^{24}$ & $2^{13.42}$ & $33920$ \\  \hline
\end{tabular} 
\caption{Complexity estimates of \Cref{alg:Keyextraction3} without $\mathcal{O}(\cdot)$ notation,  in the case when one uses (\ref{eq:recover}) for the interpolation of $G=\lambda\cdot H^{(j)}$ ($\lambda\in \mathbb{F}^n_2\setminus \{\textbf{0}_n\}$). The parameter $T$ denotes the complexity of computing the sum in (\ref{eq:recover}) with $\sum^{d-(\mu+1)}_{j=0}\binom{n}{j}$ addends. Also, $Comp_{Int}$ denotes the processing complexity of underlying interpolation.}
\label{tab4}
\end{table}
%
%

\newpage

\section{Conclusions}\label{sec:OP}

In this paper we show that the pseudo-random function Farfalle admits an application of Simon's algorithm in various settings. Several scenarios have been shown by \textbf{Constructions 1} and \textbf{2}, where much more similar combinations is clearly possible. Based on the provided constructions, we show that forgery attacks are possible to mount on Farfalle-SAE and SIV modes, as well as a construction of a quantum distinguisher for the Farfalle-WBC mode.
 The presented attacks indicate that the main weakness of Farfalle is actually the one which may potentially admit higher order differential attacks, as discussed in \cite[Section 8]{Farfale}. In context of the Kravatte instance (which is based on Keccak-$p$ permutation) and Xoofff instance (based on Xoodoo permutation), we note that their authors do not claim the quantum resistance against the attacker who can make quantum superposition queries (the \textbf{Q2} attack model). At the end, we show that one can extract a secret round key by applying two different interpolation formulas by using a reasonable amount of different periods obtained by applying Simon's or Simon-Grover algorithm to reduced-round versions (Generalized) Feistel networks in many recent papers. Especially in the case of a vector space representation of inner functions, we derive certain improvements based on trade-offs which regard different underlying parameters. In general, our methods for round-key extraction show that the existing attacks on GFNs do not only provide efficient quantum distinguishers (excluding Grover's search of certain round keys), but also one is able to derive some information related to the secret (round) key just by considering obtained periods.
 \\\\
\noindent
{\large \bf Acknowledgment:} Samir Hodžić is supported by the Slovenian Research Agency (research program P1-0404 and research projects J1-4084 and N1-1059).
\\\\

\bibliographystyle{splncs04}


\end{document}

%% file: QuantumGate.tex
\begin{tikzpicture}[thick,scale=0.8, every node/.style={scale=0.8}]

\node (xi) at (-.25,0) {$|x\rangle$};
\node[below of = xi, distance = 1cm] (yi) {$|y\rangle$};

\node[draw,minimum size = 1.5cm] (of) at (1.5,-0.5) {$\mathcal{U}_f$};

\node[align = left,right] (xo) at (3,0) {$|x\rangle$};
\node[align = left,right] (yo) at (3,-1) {$|y\oplus f(x)\rangle$};

\draw (xi) -- (0.75,0);
\draw (yi) -- (0.75,-1);

\draw (xo) -- (2.25,0);
\draw (yo) -- (2.25,-1);

\end{tikzpicture}

%% file: farfalle.tikz
\begin{tikzpicture}[
    thick,
    XOR/.style={
        thick,
        draw,
        circle,
        outer sep=2pt,
        append after command={
            [shorten >=2bp, shorten <=2bp]
            (\tikzlastnode.north) edge[thick] (\tikzlastnode.south)
            (\tikzlastnode.east) edge[thick] (\tikzlastnode.west)
        },
    },
    perm/.style= {
        draw,
        rectangle,
        minimum height=0.66cm,
        minimum width=0.66cm,
        inner sep=0pt,
    },
    perm/.default=25,
    arrow/.style= {
        ->
    },
    revarrow/.style= {
        <-
    },
    roll/.style n args={1}{
        minimum height=0.66cm,
        minimum width=0.66cm,
        inner sep=0pt,
        append after command={\pgfextra{\let\mainnode=\tikzlastnode}
        node[inner sep=0, anchor=south west] at (\mainnode.south east) {#1}},
    },
    state/.style={
        matrix of nodes,
        column sep=-\pgflinewidth,
        row sep=-\pgflinewidth,
        nodes={
            rectangle,
            draw=black,
            minimum height=0.6em,
            align=center,
            text width=0.6em,
            inner sep=0pt,
            outer sep=0pt,
        }
    },
    state/.default=2em,
    dot/.style={
        fill,
        shape=circle,
        minimum size=4pt,
        inner sep=0pt,
    },
]

\node[anchor=west] (K) at (-2.5,0) {$K||10^*$};
\node[perm] (pb) at (0,0)  {$p_b$};
\draw[arrow] (K) -- (pb);
\node[roll] (rk) at (4, -1) {\tiny $i+2$};
\circlearrow{rk};

\node[XOR] (bo) at (2, -4.5) {};
\node[perm] (pd) at (4, -4.5) {$p_d$};
\draw[arrow] (bo) -- (pd);
\draw (pd.east) -- (5, -4.5);

\foreach \xs/\ys/\yl/\xla/\yla in {0/0/-2/0/0, 1/1/-4/1/1, i/j/-7/i/j}{
    \pgfmathsetmacro{\yh}{\yl+1};
    \node[perm] at (0,\yl) (pc\xs) {$p_c$};
    \node[roll] at (-1, \yh) (rc\xs) {\ifthenelse{\equal{\xs}{i}}{}{}$\xla$};
    \circlearrow{rc\xs};
    \node at (0, \yh) (k\xs) {$k$};
    \draw[arrow] (k\xs) -- (rc\xs);
    \node[XOR] at (-1, \yl) (x\xs) {};
    \node[anchor=west] at (-2.75, \yl) (m\xs) {$m_{\xla}$};
    \draw[arrow] (m\xs) -- (x\xs);
    \draw[arrow] (rc\xs) -- (x\xs);
    \draw[arrow] (x\xs) -- (pc\xs);

    \draw[arrow] (pc\xs.east) ..controls ++(0: 1.2cm) ..  (bo);

    \node[roll] at (6.25, \yl) (re\xs) {\ifthenelse{\equal{\xs}{i}}{}{}$\yla$};
    \circlearrow{re\xs};
    \node[perm] at (8, \yl) (pe\xs) {$p_e$};
    \node at (9, \yh) (kp\xs) {$k'$};
    \draw[arrow] (k\xs) -- (rc\xs);
    \node[XOR] at (9, \yl) (xo\xs) {};
    \node at (10, \yl) (z\xs) {$z_{\yla}$};
    \draw[<-] (re\xs.west) ..controls ++(180: 0.5cm) and ++(0: 0.5cm) .. (5,-4.5);
    \draw[arrow] (re\xs) -- (pe\xs);
    \draw[arrow] (pe\xs) -- (xo\xs);
    \draw[arrow] (kp\xs) -- (xo\xs);
    \draw[arrow] (xo\xs) -- (z\xs);
}

\node (di) at (0,-5.25) {$\cdots$};
\node (do) at (6.5,-5.25) {$\cdots$};
\draw[arrow, dashed] (di.east) ..controls ++(0: 1.2cm) ..  (bo);
\draw[revarrow, dashed] (do.west) ..controls ++(180: 0.5cm) and ++(0: 0.5cm) .. (5,-4.5);
\draw[arrow] (pb) -- (k0);
\draw[arrow] (k0) -- (rk);
\draw[arrow] (rk) -- (kp0);

\end{tikzpicture}

%% file: Feistel1.tex
\begin{tikzpicture}[thick,scale=0.6, every node/.style={scale=0.6}]

	\node (x0) at (0,0) {$x_0$};
	\node (x1) [right of = x0, node distance = 4cm] {$x_1$};

	\foreach \i in {1,2,3,4}{
		\draw[thick] (0,1.75-2*\i) -- ++(0cm,-1.5cm) -- ++(4cm,-0.5cm);
		\draw[thick] (4,1.75-2*\i) -- ++(0cm,-1.5cm) -- ++(-4cm,-0.5cm);
    		\node[XOR, scale=0.8]  (xor\i) at (4,1-2*\i){};
		\node[draw, left of = xor\i, node distance = 2cm, thick,minimum width=1cm] (f\i)  {$F^{(\i)}$};
		\draw[thick,-latex] (f\i) -- (xor\i);
		\draw[thick,-latex] (0,1-2*\i) -- (f\i) ;
		\filldraw (0,1-2*\i) circle (2pt);
	}

	\node (y0) [below of = x0, node distance = 9cm] {$y_0$};
	\node (y1) [right of = y0, node distance = 4cm] {$y_1$};
	
	\foreach \z in {0,1}{\draw[thick,-latex] (4*\z,-8.25) -- (y\z); }

\end{tikzpicture}